\documentstyle[11pt,twoside,epsfig,rotate]{article}
\renewcommand{\topfraction}{1.0}

\renewcommand{\textfraction}{0.0}
\newlength{\dinwidth}
\newlength{\dinmargin}
\begin{document}
%\input{abbreviations}
%                                                    new commands
\renewcommand{\textfraction}{0.05}
\renewcommand{\topfraction}{0.95}
\newcommand{\scaption}[1]{\caption{\protect{\footnotesize  #1}}}
\newcommand{\proc}[2]{\mbox{$ #1 \rightarrow #2 $}}
\newcommand{\average}[1]{\mbox{$ \langle #1 \rangle $}}
\newcommand{\av}[1]{\mbox{$ \langle #1 \rangle $}}
%                                                    variables
\newcommand{\kjet}{\mbox{$k_{T\rm{jet}}$}}
\newcommand{\xjet}{\mbox{$x_{\rm{jet}}$}}
\newcommand{\Ejet}{\mbox{$E_{\rm{jet}}$}}
\newcommand{\thjet}{\mbox{$\theta_{\rm{jet}}$}}
\newcommand{\pjet}{\mbox{$p_{T\rm{jet}}$}}
\newcommand{\W}{\mbox{$W~$}}
\newcommand{\Q}{\mbox{$Q~$}}
\newcommand{\xB}{\mbox{$x~$}}  % Bjorken 
\newcommand{\xF}{\mbox{$x_F~$}}  % Feynman x
\newcommand{\xg}{\mbox{$x_g~$}}  % x_g
\newcommand{\y}{\mbox{$y~$}}
\newcommand{\Qsq}{\mbox{$Q^2~$}}
\newcommand{\et}{\mbox{$E_T~$}}
\newcommand{\kt}{\mbox{$k_T~$}}
\newcommand{\ptrans}{\mbox{$p_T~$}}
\newcommand{\pth}{\mbox{$p_T^h~$}}
\newcommand{\pte}{\mbox{$p_T^e~$}}
\newcommand{\ptsq}{\mbox{$p_T^{\star 2}~$}}
\newcommand{\as}{\mbox{$\alpha_s~$}}
\newcommand{\asx}{\mbox{$\alpha_s$}}
\newcommand{\ycut}{\mbox{$y_{\rm cut}~$}}
\newcommand{\gx}{\mbox{$g(x_g,Q^2)$~}}
\newcommand{\xpart}{\mbox{$x_{\rm part~}$}}
\newcommand{\mrsdm}{\mbox{${\rm MRSD}^-~$}}
\newcommand{\mrsdmp}{\mbox{${\rm MRSD}^{-'}~$}}
\newcommand{\mrsdn}{\mbox{${\rm MRSD}^0~$}}
\newcommand{\lambdams}{\mbox{$\Lambda_{\rm \bar{MS}}~$}}
%                                                    units
\newcommand{\cm}{\mbox{\rm ~cm}}
\newcommand{\GeV}{\mbox{\rm ~GeV~}}
\newcommand{\GeVx}{\rm GeV}
\newcommand{\MeV}{\mbox{\rm ~MeV~}}
\newcommand{\GeVsq}{\mbox{${\rm ~GeV}^2~$}}
\newcommand{\nb}{\mbox{${\rm ~nb}~$}}
\newcommand{\nbinv}{\mbox{${\rm ~nb^{-1}}~$}}
\newcommand{\pb}{\mbox{${\rm ~pb}~$}}
\newcommand{\pbinv}{\mbox{${\rm ~pb^{-1}}~$}}
\newcommand{\mm}{\mbox{$\cdot 10^{-2}$}}
\newcommand{\mmm}{\mbox{$\cdot 10^{-3}$}}
\newcommand{\mmmm}{\mbox{$\cdot 10^{-4}$}}
%                                                    shorthands
\newcommand{\epem}{\mbox{$e^+e^-$}}
\newcommand{\ee}{\mbox{$e^+e^-~$}}
\newcommand{\eex}{\mbox{$e^+e^-$}}
\newcommand{\pp}{\mbox{$p\bar{p}$}}
\newcommand{\qq}{\mbox{$q\bar{q}~$}}
\newcommand{\qqg}{\mbox{$q\bar{q}g~$}}
\newcommand{\qqx}{\mbox{$q\bar{q}$}}
\newcommand{\qqgx}{\mbox{$q\bar{q}g$}}
\newcommand{\ep}{\mbox{$ep$}}
\newcommand{\gsp}{\mbox{$\gamma^{*}p$}}
\newcommand{\gp}{\mbox{$\gamma p~ $}}
\newcommand{\gs}{\mbox{$\gamma^*$}}
\newcommand{\ga}{\mbox{$\gamma$}}
\newcommand{\lsim}{\raisebox{-0.5mm}{$\stackrel{<}{\scriptstyle{\sim}}$}}
\newcommand{\gsim}{\raisebox{-0.5mm}{$\stackrel{>}{\scriptstyle{\sim}}$}}
\newcommand{\PO}{I\!\!P}
\newcommand{\xpom}{x_{\PO}}
\def\figurename{{\bf Figure}}
\def\tablename{{\bf Table}}
\def\etal{{\it et al.,\ }}
\setlength{\unitlength}{1mm}
%
%==============================================================================
%\begin{document}
%%%%%%%%%%%%%%%%%%%%%%%%%%%%%%%%%%%%%%%%%%%%%%%%%%%%%%%%%%%%%%%%%%%%%%%%%%
% Title Page
\begin{titlepage}
\begin{flushleft}
{\tt DESY 97-210}\hfill {\tt ISSN 04118-9833} \\
\end{flushleft}
\vspace*{3.0cm}
\begin{center}\begin{LARGE}
\bf{Thrust Jet Analysis of \\
Deep--Inelastic Large--Rapidity--Gap Events}\\
\vspace*{2.5cm}
H1 Collaboration \\
\vspace*{2.5cm}
\end{LARGE}
%================================abstract===================
%\vspace*{0.5cm}
{\bf Abstract}
\begin{quotation}
\noindent
A thrust analysis of 
Large-Rapidity-Gap events in deep-inelastic $ep$ collisions is presented, 
using data taken with the H1 detector at HERA in 1994.
% with 
%an integrated luminosity of $1.96\,{\rm pb^{-1}}$. 
The average thrust of the final states $X$, which 
emerge from the dissociation of virtual photons
in the range $10 < Q^2 < 100~{\rm GeV}^2$, 
grows with  hadronic mass $M_X$
%in the range from 4 to 36 GeV, 
and implies a dominant 2-jet topology.
Thrust is found to decrease with growing $P_t$,
the thrust jet momentum transverse to the photon-proton collision axis.
Distributions of $P_t^2$
% in intervals of $M_X$ from 6 to 36~GeV 
are consistent with being independent of $M_X$. 
They show a strong alignment of the thrust axis with the 
photon-proton collision axis,
and have a large high$-P_t$ tail.  
The  correlation of thrust with $M_X$ is similar to that 
in \ee annihilation at $\sqrt s_{ee}=M_X$, 
but with lower values of thrust in the $ep$ data.
The data cannot be described by interpreting  the dissociated system 
$X$ as a \qq state 
but inclusion of a substantial fraction of \qqg parton configurations 
leads naturally to the observed properties.
The soft colour exchange interaction model 
does not describe the data.

\end{quotation}
\vspace*{2.0cm}
{\it Submitted to Zeitschrift f\"ur Physik $\mbox{\boldmath $C$}$}  \\
\vfill
\cleardoublepage
\end{center}
\end{titlepage}

%==============================================================================

\vfill
\clearpage
\begin{sloppypar}
\noindent
%   H1AUTS  Author list by names, no. of authors  380
%           status: 03/06/97   10.02.23
\begin{Large}
\begin{center}
H1 Collaboration \\
\end{center}
\vspace*{0.5cm}
\end{Large}
\noindent
 C.~Adloff$^{35}$,                %WUPP-ST                  Adloff              
 S.~Aid$^{13}$,                   %HAM2-LEFT    8/96        Aid                 
 M.~Anderson$^{23}$,              %MANC-ST  10/95           Anderson            
 V.~Andreev$^{26}$,               %LPI -PD                  Andreev             
 B.~Andrieu$^{29}$,               %ECPL-PD                  Andrieu             
 V.~Arkadov$^{36}$,               %ZEUT-ST    10/96         Arkadov             
 C.~Arndt$^{11}$,                 %DESY-ST   1/96           Arndt               
 I.~Ayyaz$^{30}$,                 %PARI-ST       5/96       Ayyaz               
 A.~Babaev$^{25}$,                %ITEP-PD                  Babaev              
 J.~B\"ahr$^{36}$
% ...
%ZUER-ST                  ZurNedden           
 \\
\bigskip 
 
\noindent
{\footnotesize{%     H1 Institutes as appearing on publications
 $ ^1$ I. Physikalisches Institut der RWTH, Aachen, Germany$^ a$ \\
 $ ^2$ III. Physikalisches Institut der RWTH, Aachen, Germany$^ a$ \\
% ...
$ ^5$ Rutherford Appleton Laboratory, Chilton, Didcot, UK$^ b$ \\
% ...
$ ^{11}$ DESY, Hamburg, Germany$^a$ \\
% ...
$ ^{13}$ II. Institut f\"ur Experimentalphysik, Universit\"at Hamburg,
                                     Hamburg, Germany$^ a$  \\
% ...
$ ^{23}$ Physics Department, University of Manchester,
                                          Manchester, UK$^ b$\\
% ...
$ ^{25}$ Institute for Theoretical and Experimental Physics,
                                                 Moscow, Russia \\
 $ ^{26}$ Lebedev Physical Institute, Moscow, Russia$^ f$ \\
% ...
$ ^{29}$ LPNHE, Ecole Polytechnique, IN2P3-CNRS,
                             Palaiseau, France \\
 $ ^{30}$ LPNHE, Universit\'{e}s Paris VI and VII, IN2P3-CNRS,
                              Paris, France \\
% ...
$ ^{35}$ Fachbereich Physik, Bergische Universit\"at Gesamthochschule
               Wuppertal, Wuppertal, Germany$^ a$ \\
 $ ^{36}$ DESY, Institut f\"ur Hochenergiephysik,
                              Zeuthen, Germany$^ a$\\
% ...
$ ^a$ Supported by the Bundesministerium f\"ur Bildung, Wissenschaft,
        Forschung und Technologie, FRG,
        under contract numbers 6AC17P, 6AC47P, 6DO57I, 6HH17P, 6HH27I,
        6HD17I, 6HD27I, 6KI17P, 6MP17I, and 6WT87P \\
 $ ^b$ Supported by the UK Particle Physics and Astronomy Research
       Council, and formerly by the UK Science and Engineering Research
       Council \\
% ...
$ ^e$ Supported in part by USDOE grant DE~F603~91ER40674 \\
 $ ^f$ Supported by the Deutsche Forschungsgemeinschaft \\
 $ ^g$ Supported by the Swedish Natural Science Research Council \\
 $ ^h$ Supported by GA \v{C}R  grant no. 202/96/0214,
       GA AV \v{C}R  grant no. A1010619 and GA UK  grant no. 177 \\
% ...
}}
 
\end{sloppypar}

%========================== INTRODUCTION  ======================
\vfill
\newpage
\section{Introduction}
The Large-Rapidity-Gap (LRG) events observed 
at the \ep ~collider HERA~\cite{ZEUSLRG,H1LRG} are 
attributed mainly to diffraction~\cite{H1POM,ZEUSPOM}. 
The kinematics of these events can be discussed quite generally 
in terms of the process $ep \rightarrow e'XY$ (see Fig.~1), 
%
%
%--- Figure 1 Gap definition 
%
\begin{figure}[ht] \unitlength 1mm
\label{figure:evtdefinition}
%%\begin{picture}(50,50)
%%  \put(30,50){\epsfig{file=wsfig1.eps,height=100mm,width=45mm,angle=-90}}
%%\end{picture}
\begin{center}
{\hspace{2cm}\epsfig{file=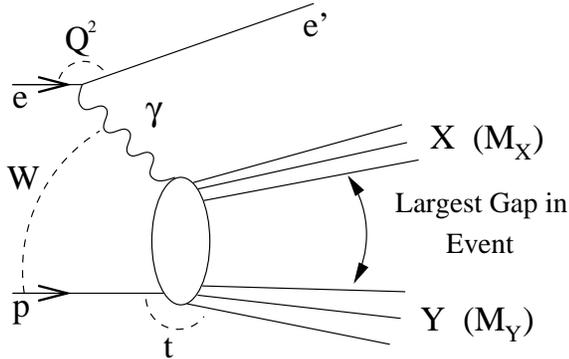,width=0.3\textwidth,angle=-90}}
\end{center}
\caption
{\small
Illustration of the process, $ep \rightarrow e'XY$, in which the hadron systems
$X$ and $Y$ are separated by the largest rapidity gap in the event. 
$W$ is the invariant mass of the colliding virtual photon - beam proton 
system. 
$-Q^2$ is the square of the 4-momentum transfer at the ($e, e'$) vertex,
and $t$ at the ($p, Y$) vertex. 
$M_{X}$ and $M_{Y}$ are the masses of systems $X$ and $Y$, respectively.
}
\end{figure}
where the two hadronic systems $X$ and $Y$ are separated 
by the largest rapidity gap in the event.
In the H1 experiment, events are selected 
with no activity over a particular large pseudorapidity range
adjacent to the outgoing proton beam (forward) direction. 
In these events the system $Y$ remains as a proton,
or is excited to a low mass baryonic system 
by a colourless exchange 
with a small transverse component in the momentum transfer~\footnote 
{Here, and throughout this paper, the term pomeron and the symbol $\PO$ 
are used for the colourless exchanged object independently of its nature,
unless the difference between a pomeron and a subleading reggeon exchange 
is significant in the context.}
$\PO = p - Y$.
The incoming photon, ranging from very low to high virtuality $Q^2$, can be 
excited~\cite{GOOD} 
into a vector meson ($\gamma^{(*)} + p \rightarrow V + Y$), 
or can dissociate 
into a high mass state $X$ ($\gamma^{(*)} + p \rightarrow X + Y$).

In this paper, the topological structure of final states, which emerge
from the dissociation of highly virtual photons,
is investigated.
In the photon dissociation (PD) picture  this process can be related to
the fluctuation of the virtual photon into 
partonic states~\cite{AJM,FOCK,BUCHGAP,BUCHJET},
followed by parton-proton quasi-elastic diffractive scattering, 
as shown in Fig.~2.
In the proton infinite momentum frame (IMF) picture
the diagrams~\footnote
{The names for the respective processes as given in the caption
are appropriate for the IMF interpretation of these diagrams.
Because the two interpretations of the diagrams in Fig.~2 are
equivalent, these names are used 
also when referring to each of Fig~2a through 2d in the PD interpretation.}
in Fig.~2 can also be described in terms 
of deep-inelastic scattering (DIS) off an exchanged object 
with partonic content~\cite{ingel_schlein,H1POM,ZEUSPOM}. 

%
%--- Figures 2 Monte Carlo diagrams
%
\begin{figure}[ht] \unitlength 1mm
%\begin{picture}(50,50)
%\begin{picture}(200,100)

\begin{center}
\begin{tabular}{cc}
\psfig{file=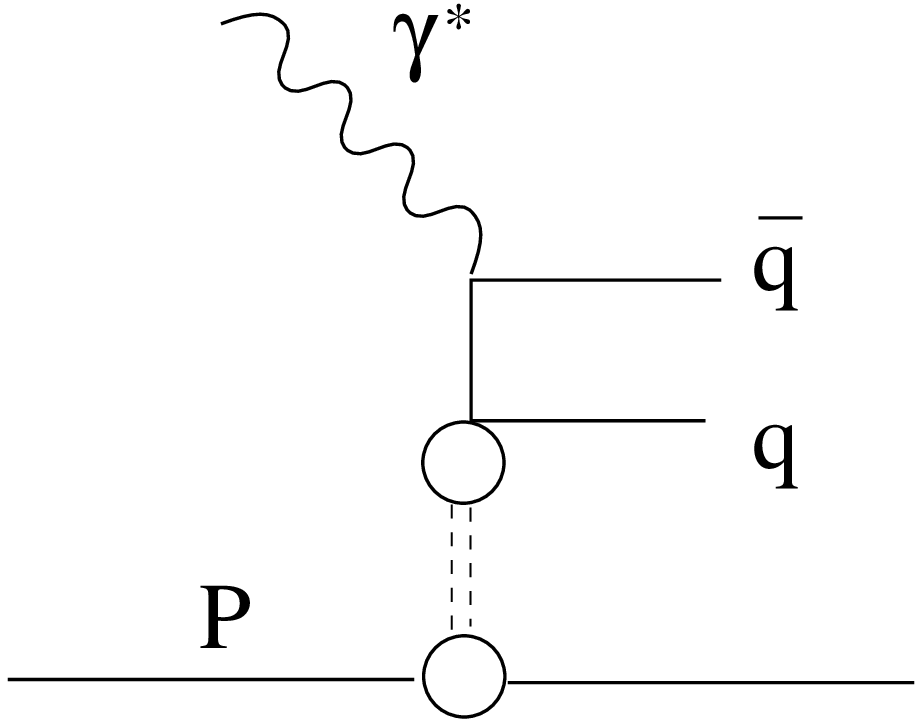,width=3cm,angle=-90}&
\psfig{file=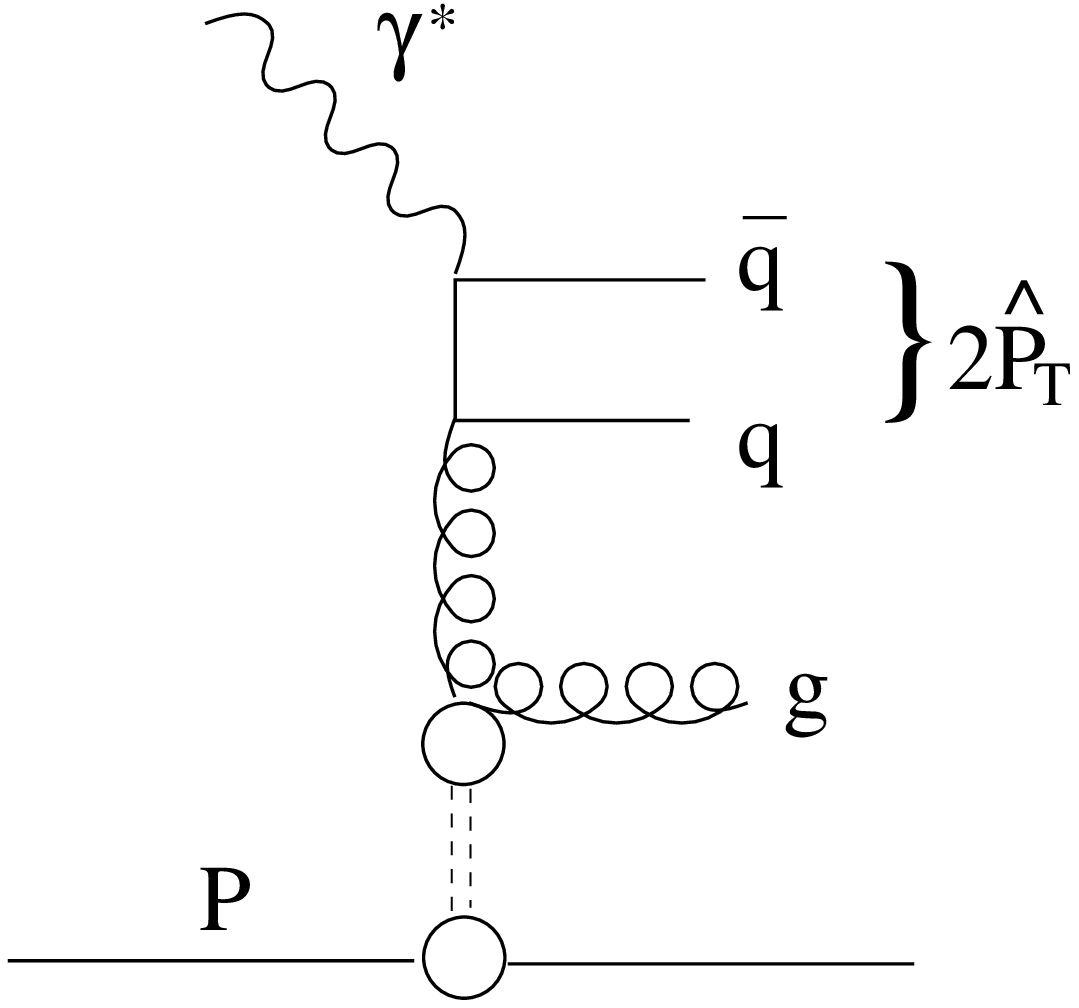,width=4cm,angle=-90}\\
(a)&(b)\\

\psfig{file=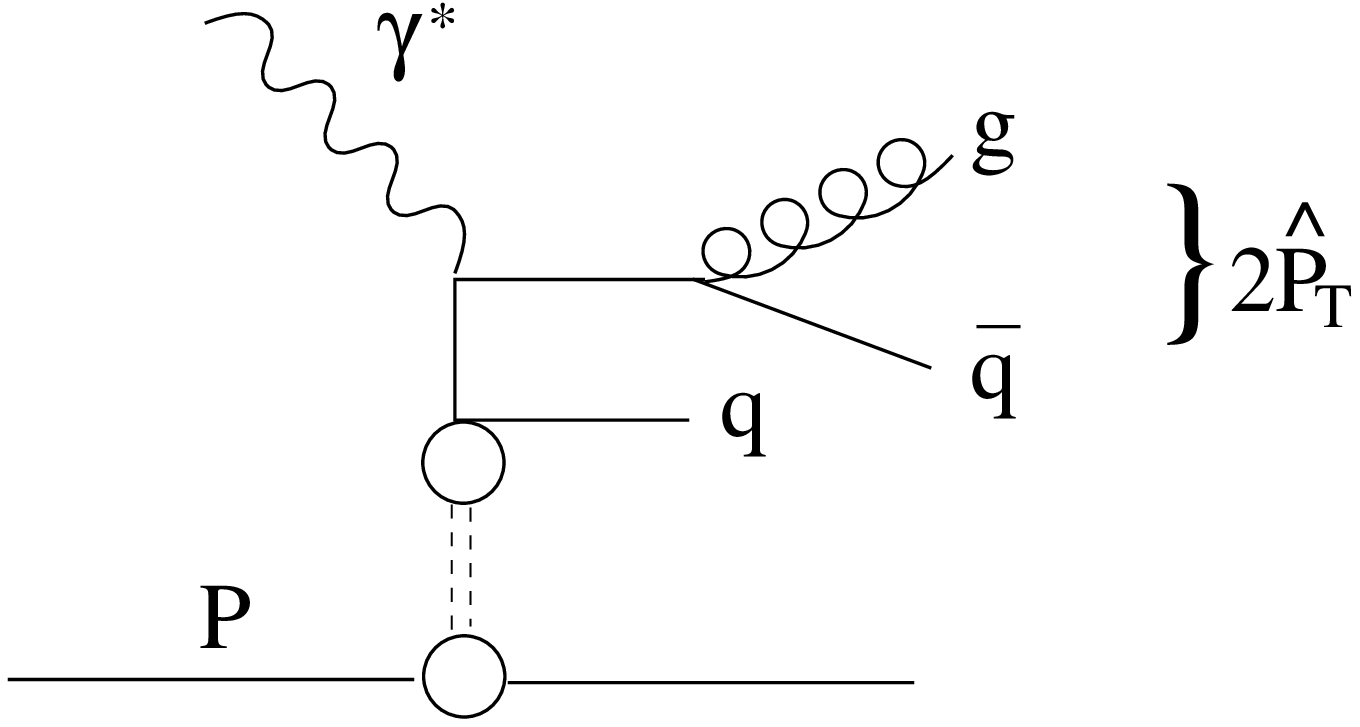,width=3cm,angle=-90}&
\psfig{file=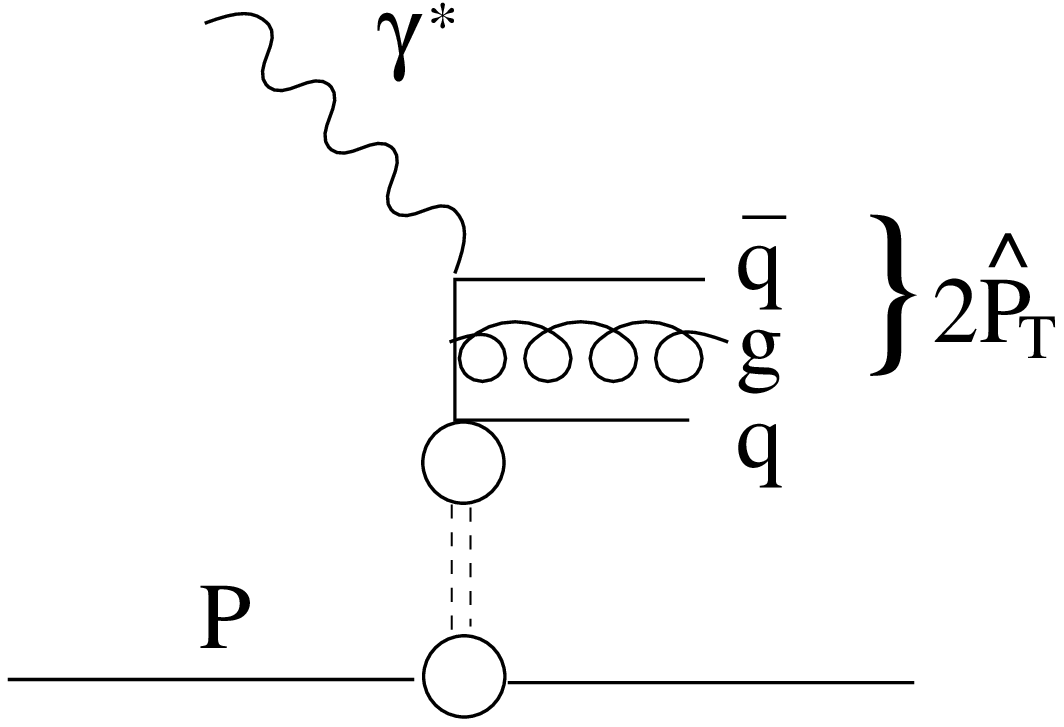,width=3cm,angle=-90}\\
(c)&(d)\\
\end{tabular}
\end{center} 
%\caption 
{{\bf Figure~2:} \small Diagrams of parton processes for LRG events:
  Born term diagram (a), 
%(b-d) for \qqg Fock states of the photon, with 
  boson-gluon fusion (BGF) (b), QCD-Compton (QCD-C) (c, d).
   The $q \leftrightarrow \bar{q}$ 
   exchanged configurations are implied.
}
\label{diagram:qqbar}
\label{diagram:qqbarg}
\end{figure}
%
%

%. ~\cite{FOCK}. 

The simplest PD process was first discussed by Bjorken and Kogut~\cite{AJM},
in terms of the quark-parton model in the form of the Aligned Jet Model (AJM). 
Here the photon fluctuates into a quark anti-quark pair
long before it actually collides with the proton.

Quark anti-quark (\qq) configurations 
with large parton momenta $k_t$ transverse to the 
collision axis are predicted to have a reduced cross section in comparison 
with the expectation for a hypothetical collision 
of two single quarks with the proton. 
This suppression of large $k_t$ has been explained in terms of 
 ``colour transparency''~\cite{COLOR_TRANS}, 
since the physical transverse separation of the \qq state is small on average,
leading to a small colour dipole moment, or equivalently to mutual screening.
However, $\qq$ configurations with  a small transverse  momentum 
difference do not lead to such screening effects
and hence are not suppressed in their interaction with the proton.
In the rest frame of $X$, the AJM picture 
thus leads to a back-to-back 2-jet configuration 
with small jet transverse momenta $P_t$
 relative to the incoming proton beam direction,
typical of soft interactions.

The QCD extension of this simple AJM model 
with gluons radiated by the \qq system
was proposed in~\cite{STRIKFURT},  
and has been studied further by several authors~\cite{FOCK},
in particular by invoking 
a soft colour interaction (SCI) process~\cite{BUCHGAP,BUCHJET}. 
In the IMF view this corresponds to a QCD evolution of the 
structure function of the diffractive exchange.

Two further and completely different approaches have been used to explain 
LRG event production.
In one of them the LRG events are the result of
normal DIS scattering
with a subsequent SCI process~\cite{SCINGEL}, 
generating a large rapidity gap.
In the other one 
it is proposed that the virtual photon fluctuates according to  
the Generalized Vector Meson Dominance Model
(GVDM)~\cite{GVDM}
into virtual vector mesons which then undergo  diffractive
scattering off the proton~\cite{STODOLSKI,SCHILDA}.

All the above approaches are phenomenological and one can alter parameters
to describe 
the inclusive production cross section for LRG events. 
They lead, however, to distinctive features in the topology 
of the hadronic final state.

Topological studies were instrumental in analysing
the transition of 
timelike photons into partons in \ee annihilation,
even at energies $\sqrt{s_{ee}}$ which were 
too low to resolve 3-jet from 2-jet events~\cite{PLUTO}.
The contribution of higher order parton configurations 
like \qqg, relative to \qq, 
was determined without applying jet algorithms,
by measuring the deviation  from colinear momentum flow
using the event-shape variable 
thrust $T$~\cite{THRUST64,THRUST77}.
Data on the evolution 
of thrust with $\sqrt{s_{ee}}$ also allowed 
a separation of hadronization from perturbative QCD effects to be made.

In this analysis, thrust 
is used to study topological features of the hadronic final state 
in \gs ~dissociation in a kinematic regime where diffractive 
contributions are dominant~\cite{F2D3_new}. 
Here the final state partonic structure and the related hadronization effects
may be different from those in \ee annihilation events.
The correlation of thrust with $M_X$
may be exploited to separate these two contributions. 
In addition, the alignment, i.e. the correlation of the thrust direction
with the incoming proton direction, can be investigated by studying 
the $P_t$ distribution of thrust jets with respect to the proton direction
and can be used 
as a further tool for the analysis of the dynamics of LRG events.

Because of a  forward rapidity gap requirement, the hadronic final states
of accepted events
are well contained in the central detectors of H1. 
This provides measurement conditions similar
to those in  \ee annihilation experiments at symmetric \ee colliders.
No jet energy threshold is required 
in the thrust analysis and 
all events in the $M_X$ range under study are included.
An event shape analysis of LRG events has been reported recently 
by the ZEUS collaboration~\cite{ZEUSSPH}.
%Large thrust jet $P_t$ values can be caused by
%large transverse momenta between \qq 
%in the Born term configuration
%- coming from the intrinsic $k_t$ at the $\gs \rightarrow \qq$ vertex 
%or from the interaction with the proton -
%or by higher order parton processes,
%i.e. mainly by 3-parton configurations. 
%Large $P_t$ caused by \qqg and higher order processes must be correlated  
%with reduced values for thrust, 
%while large $P_t$ caused by the \qq configuration
%leaves thrust and $P_t$ uncorrelated.
%A measurement of this correlation thus provides direct information 
%on the parton level process.

%=================================  DETECTOR DESCRIPTION  ===================

\section{ Detector description}
\label{sec:H1exp}
In the following, only  detector components relevant for
this analysis are  reviewed. A detailed description of the H1 detector 
can be found elsewhere~\cite{H1DET}.

The ``backward'' electromagnetic calorimeter (BEMC) has full azimuthal coverage,  
and extends over the range
$151^\circ < \theta < 176^\circ$,
where $\theta$ is the polar angle with respect to 
the proton beam direction, as seen 
from the nominal beam collision point. 
The BEMC was used to trigger on and measure the energy 
of the scattered electron in DIS processes. 
The electromagnetic energy resolution is  
$\sigma_{E}/E = 0.10/\sqrt{E\;[\rm GeV]} \oplus 0.42/E[\GeV] \oplus 0.03$
~\cite{H1BEMC},
while the absolute electromagnetic energy scale 
is known with an accuracy of 1\%.
The BEMC hadronic energy scale is known to a precision of 20\%.
The backward proportional chamber (BPC), 
located in front of the BEMC detector,
has an acceptance of $155.5^\circ < \theta < 174.5^\circ$ 
and,
in conjunction with the interaction vertex, 
 measures the direction of the scattered electron 
with a precision of 1 mrad.
These ``backward'' detectors accept DIS processes with  $Q^2$ values ranging 
from 5 to 120~GeV$^2$.
The liquid argon (LAr) calorimeter extends over the polar angular range   
$4^\circ  < \theta < 154^\circ$ with full azimuthal coverage.
The electromagnetic energy resolution is 
 $\sigma_{E}/E\approx~0.11/\sqrt{E\;[\rm GeV]}~\oplus~0.01$,
while the hadronic energy resolution is 
 $\sigma_{E}/E\approx~0.50/\sqrt{E\;[\rm GeV]}~\oplus~0.02$
as determined in test beams.
A study of the transverse momentum  balance between 
the hadronic final state and
the scattered electron, $P_{t,h} - P_{t,e^{'}}$, 
has shown that the absolute hadronic energy scale
is known to an accuracy of $4\%$.

The LAr calorimeter is surrounded by a superconducting solenoid providing a
uniform
magnetic field of $1.15$ T parallel to the beam axis in the tracking region.
The track reconstruction is based on the information from the 
central jet chamber (CJC), the $z$-drift chambers and the forward tracker. 
These detectors cover a 
polar angular range of $5^\circ < \theta < 155^\circ$.
% and provide, with a maximum of 56 points per track, a momentum resolution of
%$\delta p_{t}/p_{t}\leq 0.01~p_{t}/$\rm GeV. 
\par
 Forward energy deposits at small angles are observed in several detectors 
near the outgoing proton beam direction.
Particles reach these detectors both 
directly from the interaction point, and indirectly
as a result of secondary scattering with the beam pipe wall
or adjacent material such as collimators.
These detectors are thus sensitive to particles well outside
their nominal geometrical acceptances. 
The liquid argon calorimeter is sensitive to particles with 
pseudorapidities $\eta = -\ln \tan \theta/2$ up to $\eta ~\simeq~5.5$. 
A copper-silicon sandwich
calorimeter (PLUG) allows energy measurements to be made over the range
$3.5<\eta<5.5$. 
The three double layers of drift chambers of the forward muon
detector (FMD) are sensitive to particles produced at pseudorapidities
$5.0<\eta<6.5$. The proton remnant tagger (PRT), consisting of seven double
layers of lead/scintillators, located 24~m from the interaction point,
 covers the region $6.0<\eta<7.5$.

These detectors overlap considerably in their rapidity coverage,
thereby allowing intercalibration of their efficiencies.
\\

\par 
%======================  MODELS =======================================

\section{Physics models for LRG event production}
In the PD picture
the photon fluctuates into a \qqx, \qqgx, or states with more partons,
where at least one of the partons is virtual. 
A virtual parton, if it has a low 
$k_t$ relative to the \gsp ~collision axis, 
can scatter diffractively (quasi-elastically) with the proton yielding 
a real parton. 
In the \qq configuration (Fig.~2a) this gives  
an aligned 2-jet event,
and at least one aligned jet in the 3-parton configurations 
(Figs.~2b, c and d).
It is expected  qualitatively that the \qqg configuration
has a larger cross section than the \qq configuration
for diffractive scattering off the proton~\cite{BUCHJET}.
%Calculations for topological properties of such configurations
%are in progress in the PD picture~\cite{DERMOTT}.

In the IMF picture the diagrams in Fig.~2 
are interpreted as the DIS probing a colourless exchange object. 
This is implemented in the RAPGAP~\cite{RG} Monte Carlo (MC) program.
The RAPGAP model employs deep-inelastic electron scattering
off pomerons~\cite{ingel_schlein} and off (subleading) reggeons coupling to 
the proton. 
The pomeron is ascribed a quark and a gluon content.
The ratios of the BGF to the Born term and to the QCD-C contributions 
(see Fig.~2) depend on
the gluon and quark contents of the exchanged objects. 
These have been determined from a  QCD analysis with 
DGLAP evolution~\cite{DGLAP} 
of the diffractive structure function measured by 
the H1 collaboration~\cite{F2D3_H1}, 
with the result that most of the pomeron momentum is carried 
by gluons.
For the reggeon, the quark and gluon content is taken to be that of the
pion~\cite{PIONSF}.

The RAPGAP model is used in two modes.
The matrix element mode (ME) includes 
the BGF and the QCD-C processes as first order 
perturbative QCD (pQCD) matrix elements,
with propagators for the parton mediating the hard process,
while the diagram in Fig.~2a (Diagram~2a), 
is treated as DIS off a quark in the pomeron.
The ARIADNE~\cite{ARIA} mode (AR) also 
treats the BGF process as a matrix element,
but treats Diagrams 2c and 2d as colour dipole radiation 
from the \qq pair of Diagram 2a. 
A cut-off in terms of the square of the transverse momentum 
of the hard sub-process, $\hat p^2_t ~\ge ~\hat p^2_{t,min} (=2$~GeV$^2)$
is applied to keep the contribution from diagrams treated as 
pQCD matrix elements
below  the measured total cross section at each $(M_X, Q^2 , x)$ point.
With this choice~\cite{RG} of cut-off, 
the corresponding diagrams of Fig.~2 contribute 
about $50\%$ of the total diffractive cross section within RAPGAP.
The remainder is generated with the process of Diagram 2a. 
The cut-off value can be raised, 
but it cannot be lowered without the need to readjust other parameters
influencing the total cross section.
% and topological distributions.
%The ratio of \qqg to \qq events,which
The topological properties, like thrust or thrust jet $P_t$,
of the simulated events
are sensitive to  the fraction of matrix element events
and depend on the value chosen for $\hat p_{t,min}$.

In the RAPGAP ME mode 
higher order QCD effects are included by means of leading-log parton showers 
for all final state partons.
Predictions of this mode are labelled as `RG ME+PS' in the following.
In the AR mode higher orders are included 
as additional gluon radiation 
according to the Colour Dipole Model~\cite{CD} (labelled as `RG AR+CD').
In both modes the hadronization  is as given
by the Lund string fragmentation model~\cite{STRING}.

In generating events corresponding to Diagram 2a 
the \qq system is taken to be aligned exactly with the
$\gs \PO$ axis, i.e. without intrinsic $k_t$.
The virtuality of the quark mediating the interaction 
between the photon and the pomeron is thus 
always taken as the smallest value which is kinematically allowed.
These events can be considered as representing the 
simple AJM, 
which differs from the RAPGAP simulation of Diagram~2a
by having no QCD dipole radiation, 
and by neglecting transverse momentum transfer to the proton. 
The corresponding predictions 
are labelled as  `RG q$\bar{\rm q}$'.

The non-diffractive \ep~DIS plus SCI picture is implemented in  
the LEPTO event generator~\cite{LEPTO6.4}. 
This employs the same hard processes as shown in Fig.~2,
but now off the proton.
The ratio of the generated \qqg to \qq events is determined
by an infrared cut-off~\footnote{The infrared safety scheme in
LEPTO
avoids a sharp cut-off in $\hat p_t$.} and the parton density functions 
for the proton. 
Both are adjusted to describe the total DIS event sample, but there 
is no parameter in LEPTO
to adjust specifically for the topological properties of the
LRG events.
This is different from the RAPGAP model
where the relative gluon and quark content of the pomeron
and the $\hat p_{t,min}$ value can be adjusted 
separately to describe the LRG event data.
The parameter which controls the strength of the SCI process
has no influence on the topological properties of the system $X$. 
The SCI process, and hence the LRG which separates the systems $X$ and $Y$
in Fig.~1, 
is not limited to occur between the remnant partons
of the struck proton and those emerging from the hard process. 
Thus, the final state $X$  can in general not be identified
with a specific hard process. 
%
%\bigskip\\

Predictions from the GVDM picture for final state properties 
require further specifications of the diffraction process as 
a hadronic collision.

\

%=============================== thrust analysis method ======================
\section{ Thrust jet analysis method}
In the centre of mass of a  system $X$ of $N$ particles, the thrust method 
determines the direction of the unit vector $\vec a$ 
along which the projected momentum flow 
is maximal~\cite{THRUST64}. Thrust is computed as~\cite{THRUST77}:

 $$ T = (1/\sum\limits_{i=1}^N|\vec p_i|)\cdot\max_{\vec a} 
\sum\limits_{i=1}^N|\vec p_i \cdot \vec a| $$
where $\vec p_i$ represents the momentum of particle $i$, in the rest frame 
of the $N$ particles~\footnote{
There is another definition of thrust ($T_M$)
in the literature~\cite{THRUST77}, where the normalization is not to
$\sum\limits_{i=1}^N|\vec p_i|$ as in $T$ but rather to $M_X$,
%$$ T_M = 1/M_X\cdot\max_{\vec a} \sum\limits_{i=1}^N|\vec p_i.\vec a|, $$
where $M_X$ is the invariant mass of the system $X$. 
If final states contain  massive particles, $T_M < T$. 
To compute $T_M$ accurately, 
it is necessary to identify all final state particles,
which is usually not possible experimentally.
}.

Given the thrust axis $\vec a$, the directional sense of which is arbitrary,
the $N$ particles can then be grouped uniquely 
into two subsets (thrust jets), depending on whether they
belong to the hemisphere with positive or negative momentum component along 
the thrust axis.
The summed particle momenta of hemisphere $I$ form the jet 4-momentum 
$P_i = \sum\limits_{j=1}^{N_I} p_j$ 
% and $P_2 = \sum\limits_{k=1}^{N2} p_k$ 
with $N_I$ the number of particles in hemisphere $I$, for $I=1,2$.
The two thrust jets have independent masses,
and equal but opposite 3-momenta: $|\vec P_{1,2}| = P$.  

Thrust values range from a maximum value of $T = 1$ in the case of
a 2-particle state or any colinear configuration,
to a minimum value of $T=0.5$ obtained in
an isotropic system $X$ with infinite multiplicity.
A symmetric 3-particle configuration yields a value $T = 2/3$  
and leaves the direction $\vec a$ arbitrary in the 3-particle plane,
while a non-symmetric topology gives $T > 2/3$ and a thrust 
axis pointing in the direction of the most energetic particle.
A non-symmetric 3-particle topology will thus appear as a 2-jet like
configuration with $T < 1$.
\par
In a multihadron state emerging from a partonic process, 
the two back-to-back reconstructed thrust jets are 
correlated with the hard partons in the following way.
In the case of a 2-parton configuration,
the thrust value as determined from the final state hadrons -- i.e. 
at the hadron level -- is smaller than 1, and 
%and will be reduced at hadron level to a value $T_{had} = P_{i}/P_{pa}$. 
the direction of the thrust axis remains parallel to the 
direction of the two partons to the extent that the hadrons
can be correctly assigned to `parent-partons'. 
For an underlying 3-parton system, 
%even with insufficient energy ($M_X ~\lsim~20$~GeV) 
%to form three separable jets at the hadron level,
% the direction of 
the reconstructed thrust axis 
at the hadron level is correlated with the direction of the
most energetic parton. This property has been verified 
to persist down to final state masses $M_X ~\sim~5$~GeV, 
using
\qqg events from the RAPGAP event generator~\cite{RG} with
hadronization and detector effects included in the 
event simulation.

\par
A  transverse thrust jet momentum $P_t$ can be defined
relative to a reference direction $\vec r$ as $P_t = P \cdot \sin \Theta$
with $\Theta$ the angle between $\vec P_i$ and $\vec r$.  
In this analysis, the proton beam direction transformed into
the centre of mass frame of the system $X$ is chosen for $\vec r$.
Since the 4-momentum transfer squared $t$ is small in LRG events, 
this direction is a good approximation 
to the $\gamma^* \PO$ axis. 
For $t = t_{min}$ they are identical.
Monte Carlo studies have shown that a diffractive $t - t_{min}$ distribution
with an exponential slope parameter $b = 6$~GeV$^{-2}$
leads to an average smearing of $P_t$ of less than $0.3$~GeV.
Unlike the $\gs$ direction, when calculated from the scattered electron,
the proton direction in the $X$ rest frame
is not affected by QED-radiation. 
The $\PO$ direction is not well determined experimentally.

If events with large thrust jet $P_t$ 
are dominantly caused by single gluon radiation 
from an aligned \qq state, 
then large $P_t$ will be correlated
with lower average thrust $\langle T\rangle $.
If the larger $P_t$ values were instead caused by a rotation  
of a \qq configuration to larger angles, i.e. by larger intrinsic $k_t$, 
then thrust would not change with $P_t$.
Hence,  the observation of a decrease of $\langle T\rangle $ with
increasing $P_t$ is evidence for the presence of 
\qqg contributions in the final state.

%
%============================= event selection ============================
%

\section{Event selection and final state reconstruction}
The DIS event sample used in this analysis was obtained from data collected
during the 1994 HERA operation, and corresponds to an integrated luminosity
of about
$2\,{\rm pb^{-1}}$. The event trigger required a minimum energy deposition 
in the BEMC of $4$~GeV. 
The events were selected according to the  kinematical properties
of the scattered electron~\footnote{
The data analysed here were taken with a positron beam in HERA. 
The term electron is generic.}~\cite{F2_H1}
and the hadronic final state. The following variables are used:

\begin{center}
%$Q^2 = 4 E_{e}E_{e^{'}}\cos^2\frac{\theta_{e^{'}}}{2}~~~~
%y_{e}=1-\frac{E_{e^{'}}}{E_{e}}\sin^2\frac{\theta_{e^{'}}}{2}~~~~
%x = \frac{Q^2}{ y_{e}s}$
$Q^2 = 4 E_{e}E_{e^{'}}\cos^2(\theta_{e^{'}}/2)~~~~
y_{e}=1-{E_{e^{'}}/E_{e}}\sin^2(\theta_{e^{'}}/2)~~~~
x = Q^2/(y_{e}s)$,
\end{center}
calculated from the electron beam energy $E_{e} (=27.5$~GeV),  
the energy $E_{e^{'}}$ and the polar angle $\theta_{e^{'}}$
of the scattered electron,
and $\sqrt s$, 
the invariant $ep$ mass.
The invariant mass $M_{X}$ of the hadronic system $X$ is calculated as
a geometric mean : 
\begin{equation}
M_X^2=M_{X}(e)~M_{X}(h)
\end{equation}
with
\begin{center}
$M_{X}^2(e)=[E_{e}-P_{z,e}-(E_{e^{'}}-P_{z,e^{'}})](E_{h}+ P_{z,h})- P_{x,h}^2- P_{y,h}^2 $
\end{center}
%\newline
\vspace{0.3cm}
\begin{center}
$M_{X}^2(h)=E_{h}^2- P_{z,h}^2- P_{x,h}^2- P_{y,h}^2 $
\end{center}
where 
\begin{center}
$      E_{h} =\sum_{hadrons} E_{i}~~~~~~~ P_{x(y,z),h}=\sum_{hadrons} P_{x(y,z),i} $
\end{center}
and $E_{i}$, $P_{x,i},P_{y,i},P_{z,i}$ are 
the individual energy and momentum components 
of the hadronic final state in the laboratory system.

The hadronic 4-momenta were calculated using the interaction vertex and
the energy deposit clusters in the calorimeters.
In addition the measured momentum  from each track was included 
up to a maximum of 350 MeV/c
to compensate for losses in the calorimetric measurement of charged hadrons.
This limit was determined by the requirement of a balance 
between $P_{t,e^{'}}$, the transverse momentum of the scattered electron, 
and  $P_{t,h}$, the transverse momentum of the observed hadrons. 
For the LRG events, which have only 
very small unobserved $p_t$ contributions in the forward direction,
this procedure yields
a width of 1.2~GeV in the $P_{t,h}-P_{t,e^{'}}$ distribution.
%This method of supplementing the calorimetric measurements with track
%measurements, such that the $P_{t,h}-P_{t,e^{'}}$ balance is restored,
%reduces the dependence of the systematic errors 
%on the uncertainty of
%the hadronic energy scale of the LAr calorimeter (see section~2).
%The systematic error estimates determined in the following section 
%do not include this reduction
%and  are thus a conservative.
%instead of $0.84$ as for calorimetric momenta only. 
%The balance distribution gets narrower, 
%both in data and in simulated events.

\par

The reconstruction errors for $M_X(e)$ and $M_X(h)$
are mostly independent. 
Large errors can occur because of initial state QED radiation [$M_X(e)$]
or from acceptance losses in the backward direction [$M_X(h)$].
Studies with simulated events show that using the geometric mean [equ.(1)]
reduces extreme deviations from the true $M_X$.

From the above quantities and with $q$ the 4-momentum  of \gs  ~the 
variables $\beta$ and $\xpom$ can be calculated as follows
%\begin{equation}
$$
  \beta = \frac{-q^2}{2q\cdot (p-Y)} =\frac{Q^2}{Q^2+M_X^2-t}
$$
%            \label{eq:betaQ2} 
%\end{equation}
%\begin{equation}
%{\rm and}  \,\,\,\,  
$$
\xpom =\frac{q\cdot (p-Y)}{q\cdot
    p}=\frac{Q^2+M_X^2-t}{Q^2+W^2-M_p^2} = \frac{x} {\beta}.
%           \label{eq:xpomQ2}
%\end{equation}
$$
%\begin{center}
%$\beta = \frac{Q^2}{Q^2 + M^2_{X}},
%~~~~\xpom = \frac{x}{\beta}$.
%\end{center}
In the ``proton infinite momentum frame'', the variable $\xpom$ can be interpreted
as the fraction of the proton momentum transferred  
to the photon by the exchanged $\PO$, 
while $\beta$ is the 
fraction of the $\PO$ momentum carried by the parton coupling to the photon.

The 4-momentum transfer squared, $t$, between the incident proton
and the final state $Y$ is small in LRG events
since the forward detector selection 
amounts to ${|t|} < 1$  GeV$^2$.
Therefore $t$, which is not measured,  
can be neglected in the above formulae. 

The following 
selection criteria were applied to the scattered electrons of the LRG sample: 

\begin{enumerate}
\item The electron energy was constrained  to 
 be large ($E_{e^{'}} > 14.4$~GeV)
by requiring $y_e < 0.5$. 
This keeps the photoproduction background at a negligible level, 
and removes events where the hadronic system $X$ is boosted strongly towards 
the BEMC and  the backward beam pipe hole.
\item 
The electron scattering angle was limited to  $156^\circ < \theta_{e^{'}} < 173^\circ$
to avoid events with electron energy deposition near the edges of the BEMC.
\item $Q^2$ was required to be in the range 10 to 100 GeV$^2$. 
In this $Q^2$ range there is almost constant acceptance for electrons.
\end{enumerate}
Further selections were made with quantities derived from the full final state:
\begin{enumerate}
\item 
The interaction vertex was required to be within $\pm 30$ cm 
of its nominal position and to be
reconstructed  with at least 1 track in the central jet chamber. 
\item  $\xpom<0.05$ was required to suppress non-diffractive contributions.
\item The largest rapidity gap was required to include the 
pseudorapidity range 
of the forward detectors (see section 2) 
by demanding the absence of 
significant activity in these detectors 
and, for $\eta > 3.2$, in the LAr calorimeter. 
These requirements have a high efficiency to reject
events with particles in the region
 $7.5 > \eta > 3.2$~\cite{andythesis}.
%${\rm E_{Plug} < 3\rm GeV}$,
% the number of track segments in the forward muon tracker $\leq 1$, 
%no hit in the proton tagger and no energy deposition (${\rm E<0.4\GeV}$) 
%in the LAr calorimeter for $\eta > 3.2$ (``$\eta_{max}$~cut'') 

\item  The mass $M_X$ of the hadronic final state 
was required to be bigger than 4 GeV,
the approximate limit below which the thrust axis 
ceases to be correlated with a 2-parton axis.
\item  $M_X < 36$~GeV was required to avoid events where the 
rapidity gap requirement implies an acceptance 
below $25\%$.

\end{enumerate}

With these selection criteria 6865 events were collected in the 
$M_X$ range from 4 to 36~GeV. 
They are distributed over the $M_X$ intervals  
as shown in Table~\ref{table:mxevts}.
These events cover the range $10^{-4}~< ~x~ <~3.2\cdot 10^{-2}$, 
and $30~<~W<~210$~GeV. 
%breakdown of the selected events in the 7 mass 
%intervals used for the thrust analysis, covering the mass 
%range from 4 to 36 GeV, is shown in Table \ref{tabel:mxevts}. 
\begin{table}[ht]\centering
\begin{tabular}{|c|c|c|c|c|c|c|c|} \hline
~~~~~~~~~ & \multicolumn{7}{c|}{$M_X$ intervals (GeV)}         \\   \cline{2-8}
~~~~~~~~~ & 4-6  & 6-8  & 8-11  & 11-15 & 15-19 & 19-24 & 24-36 \\  \hline
No. of events  & 1439 & 1278 & 1404  & 1130  & 711   & 502   & 401   \\  \hline
\end{tabular}
%\begin{flushleft}
\caption{\small Number of observed events per measured $M_X$ interval.}
\label{table:mxevts}
%\end{flushleft}
\end{table}
%corrected &1722$\pm$66  & 1501$\pm$61 & 1640$\pm$63 & 1466$\pm$60 & 
%           1119$\pm$59 & 944$\pm$69  & 961$\pm$96              \\
%\begin{table}[ht]\centering
%\begin{tabular}{|c|c|c|}                                           \hline
%~~~~~Variable~~~~~ &~~~~~~~~~~ Range ~~~~~~~~~~ & ~~~~~Mean value~~~~~          \\  \hline
%$Q^2$      & $10 - 100$~GeV$^2$            &  ~22~GeV$^2$       \\
%$x $   & $10^{-4} - 3.2\cdot 10^{-2}$      &  ~~$2\cdot 10^{-3}$     \\
%$\xpom$    & $10^{-4} - 5\cdot 10^{-2}$           &    0.014   \\
% $W$         & $30 - 210$~GeV                &  136~GeV          \\  \hline
%\end{tabular}
%\caption{Event sample characteristics for  $4 \leq M_X \leq 36$~GeV,
%$before acceptance corrections.}
%\label{table:evtpars}
%\end{table}
%

%
%============================  reconstruction  ================
%

\section{Corrections for resolution and acceptance}
\label{sec:resocor}

Using simulated events, the measured distributions have been corrected
for detector resolution and acceptance losses
to give cross sections in a kinematic range defined by
\begin{equation}
M_Y<1.6~{\rm {GeV}^2}, ~{\xpom}<0.05, ~{|t|}< 1~{\rm {GeV}^2},
~10<Q^2<100~{\rm {GeV}^2}~ {\rm {and}}~~y<0.5, 
\end{equation}
%$M_Y<1.6$~GeV, ${\xpom}<0.05$, ${|t|}< 1$~GeV$^2$,
%$10<Q^2<100~{\rm {GeV}^2}$ and $y<0.5$, 
for the same $M_X$ intervals as given in Table~\ref{table:mxevts}. 
Events with these $M_Y$, ${\xpom}$ and $t$ limits  
have large rapidity gaps in the detector  for kinematic reasons~\cite{H1POM}.
This cross section definition
only allows for small non-diffractive contributions,
which are typically less than $10\%$~\cite{F2D3_new},
and it avoids the need for a statistical subtraction  from the data
of a model-dependent non-diffractive background.
It also reduces the systematic error due to
model dependence of  acceptance corrections.
%To this end the  hadronic final state in each generated event is decomposed
%into two subsytems $X$ and $Y$ (Fig.~1) which are separated by the
%largest rapidity gap between 2 hadrons.
%For the definition of diffractive like events $M_X$ and $M_Y$ 
%are requested to be much smaller than the total hadronic invariant mass $W$ 
%of the system.

\par
Two different Monte Carlo event simulations
were used to correct the observed experimental distributions.
These were RAPGAP in the ME+PS and in the AR+CD mode,
but with simple (flat) quark and gluon distribution functions 
for the pomeron and without $Q^2$ evolution. 
The generated events were passed through a full detector simulation
and had the same event reconstruction as real data.
The Monte Carlo simulation of this simplified 
model described quite well the kinematic distributions of   
$Q^2, \xpom$ and $W$,
and further variables describing the detector response.
The observed $P_t$ distributions were described qualitatively. 
With an additional smooth weighting function for the generated events, 
which depends on thrust jet $P_t$ at the hadron level, 
the $P_t$ distributions in all $M_X$ intervals 
were very well approximated over the full $P_t$ range.
The measured thrust distribution
 is also well reproduced
after the weighting in $P_t$, as shown in Fig.~3.
%%%%%%%%%% replaced by Fig6,, 

%%--- Figure 3 T _ Mx correlation, comp with MC
%
\begin{figure}[h] \unitlength 1mm
\label{figure:thrust_obs}
%\begin{center}
\begin{picture}(140,120)
     \put(15,0)
{\epsfig{file=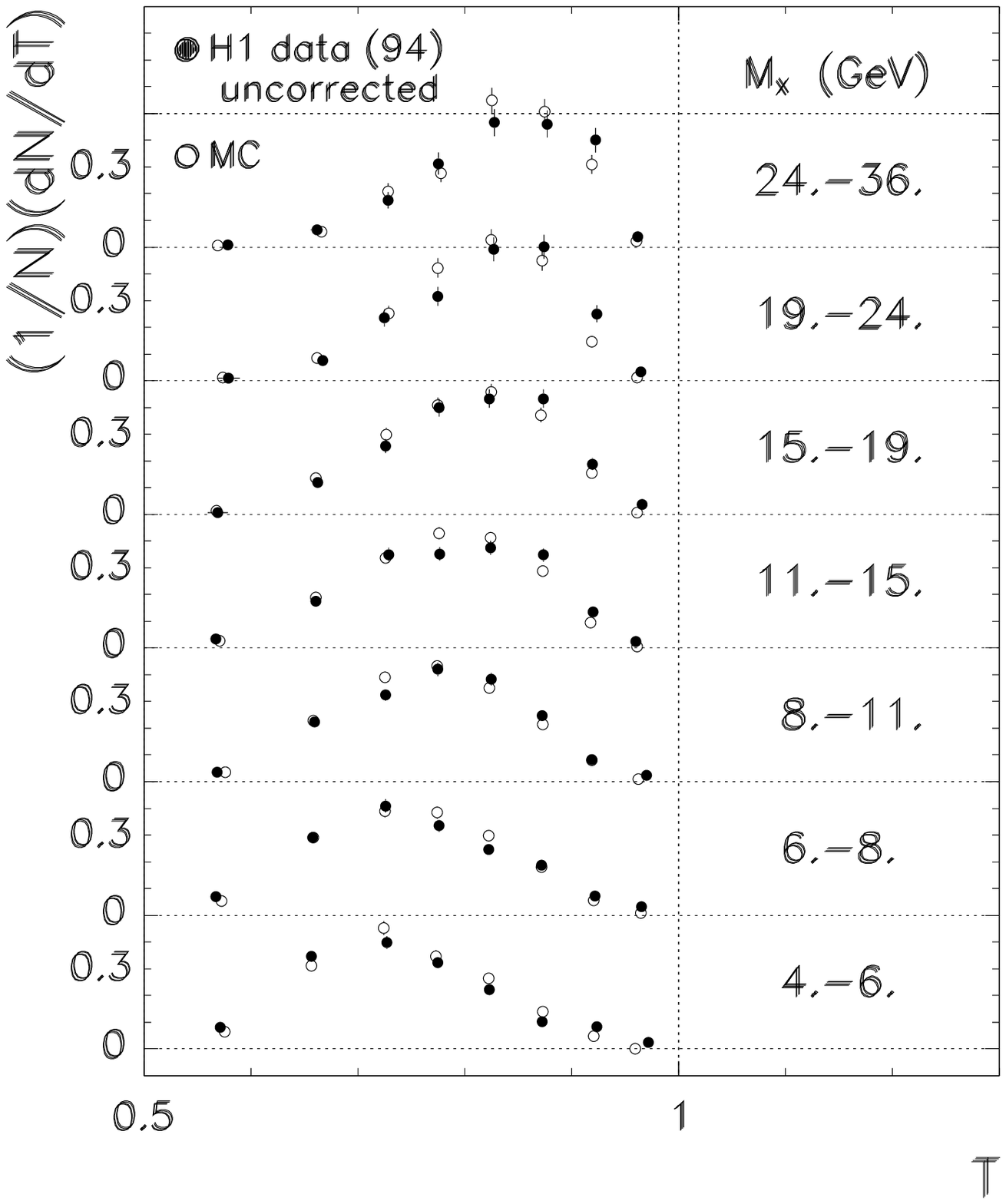,width=130mm,height=130mm}}
\end{picture}
%    \parbox[t]{1cm}{~}
%    \hspace{1cm}
%% This is the width of the text under the figure
    \parbox[t]{15.cm}
%    \parbox[t]{13.0cm}
  {{\bf Figure 3:}\small
    ~Normalized, uncorrected thrust distribution, 
    with $N$ the number of observed events, 
    for data (full points) and the simplified, 
    $P_t$-weighted MC simulation (open points), as described
    in the text.
  }
%\end{center}
%    \hspace{.5cm}
%    \parbox[t]{6cm}
%  {Figure 5b:                  
%      Blablablabla}
\end{figure}

The contents of the 2-dimensional bins 
of the experimental $(P_t^2, M_X)$ and $(T, M_X)$ distributions were multiplied
by the ratios of the numbers of generated to reconstructed and selected events 
in the corresponding intervals for the MC distributions.
%Then the 1-dimensional $P_t^2$ and $T$ distributions in each $M_X$ bin 
%were normalized.

\par  
Monte Carlo studies show that the 
thrust jet $P_t$ is resolved to better than 1 GeV.
The correction factors vary only slowly 
from bin to bin in all $P_t^2$  distributions and are large ($>2$) only for 
masses $M_X>15\GeV$ and for $P_t^2<1$~GeV$^2$.
In the thrust distributions,  
the correction factors in each mass bin 
vary less than a factor two,
except for the lowest and the highest thrust bins. 
Since those intervals contain only 
a very small fraction of the events (see Fig.~3),
 these factors have very little influence 
on the average thrust value.
The corrected average thrust is calculated from the
corrected thrust distribution.
\par
To estimate the systematic errors, the following contributions were
taken into account:
\begin{itemize}
%\hspace*{0.5cm}
%$\bullet$
%$-$  
\item uncertainty in the LAr hadronic energy scale;
%\hspace*{0.5cm} 
%$-$  
\item uncertainty in the BEMC hadronic  and electromagnetic energy scales;
%\hspace*{0.5cm}
%$-$  
\item uncertainty in the correction factors due to freedom
in the $P_t$ weighting;
%\hspace*{0.5cm}
%$-$ 
\item differences between 
event generation schemes (parton shower, colour dipole).
%\hspace*{0.5cm}
\end{itemize}
The two latter contributions 
were the largest sources of systematic error.
Their effect was determined for each data point 
from the differences between the four correction factors coming
from the two event generation schemes -
with colour dipole radiation or with parton showers - 
and each of them in the unweighted and  the weighted mode.
The unweighted $P_t$ distributions at detector level of the simulated events 
were slightly flatter than the data, 
and the weighted ones were slightly steeper.
\par

  Radiative corrections have been applied using the programs 
HERACLES~\cite{HERACLES} and RAPGAP in the matrix element and ARIADNE modes,
as described in Section~3.
The corrections  are smaller
than the statistical errors for all data points shown.

\par
The statistical, and the 
statistical and systematic errors added in quadrature,
are indicated in the figures as
the inner and the outer error bars, respectively.
Systematic and statistical errors are given 
separately in the respective tables. 

\
%
%================================ RESULTS  ==================================

\section{Results}
\subsection{Correlation of thrust
with \mbox{\boldmath$M_X, P_t$} and \mbox{\boldmath$Q^2$}}
 
In Fig.~4a thrust is shown as a function of $1/M_X$, 
averaged in seven $M_X$ intervals (see Table~\ref{table:thrust}) 
in the range 4 to 36~GeV, and 
averaged over the kinematic region of equ.(2). 
The same data are also given in Table~\ref{table:thrust}. 
%
%--- Figures 4a/4b  Thrust / 1/Mx
%
\begin{figure}[ht] \unitlength 1mm
\label{figure:thrustdata}
\begin{picture}(150,110)
       \put(-8,0){\epsfig{file=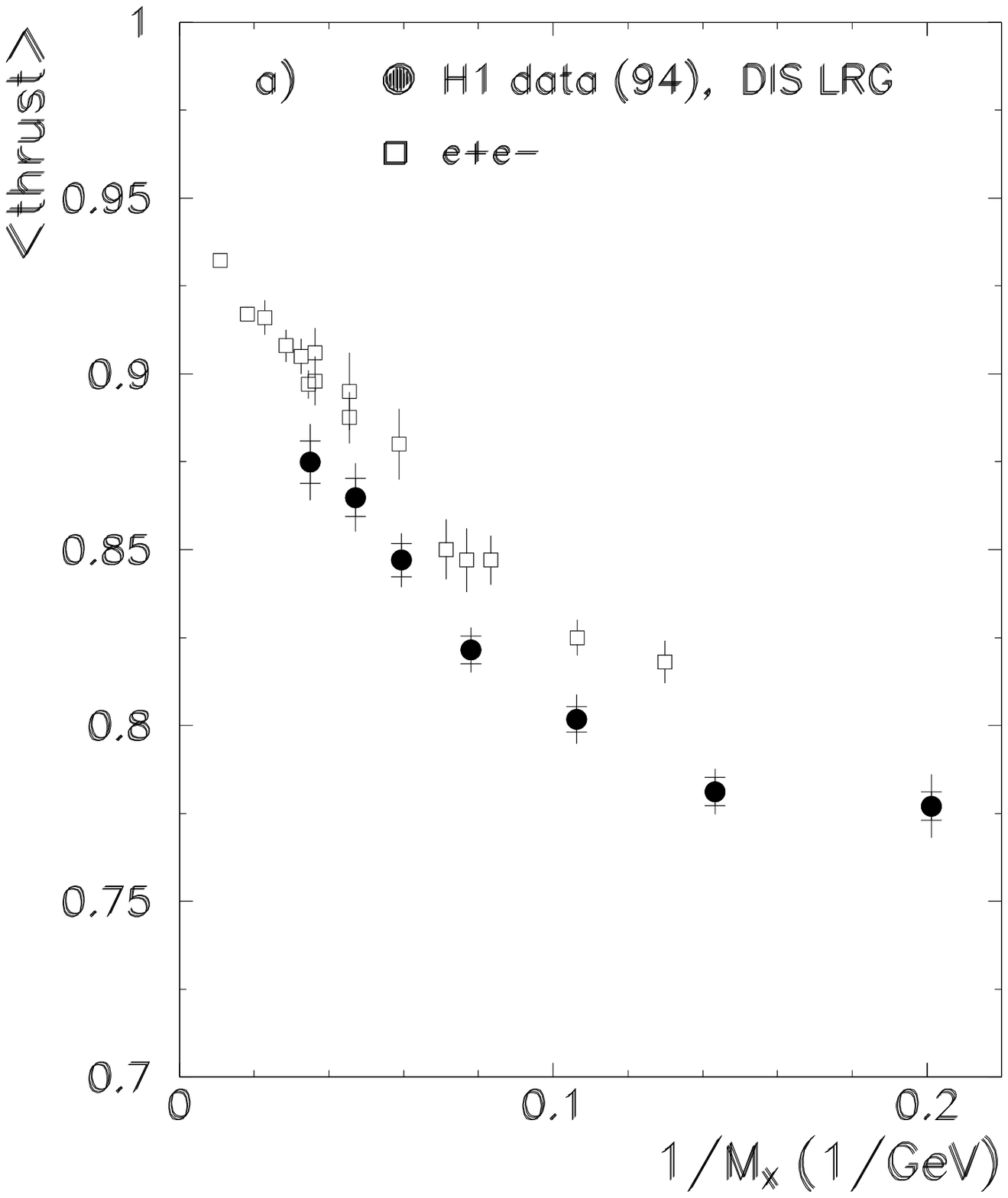,width=110mm,height=120mm}}
       \put(60,0){\epsfig{file=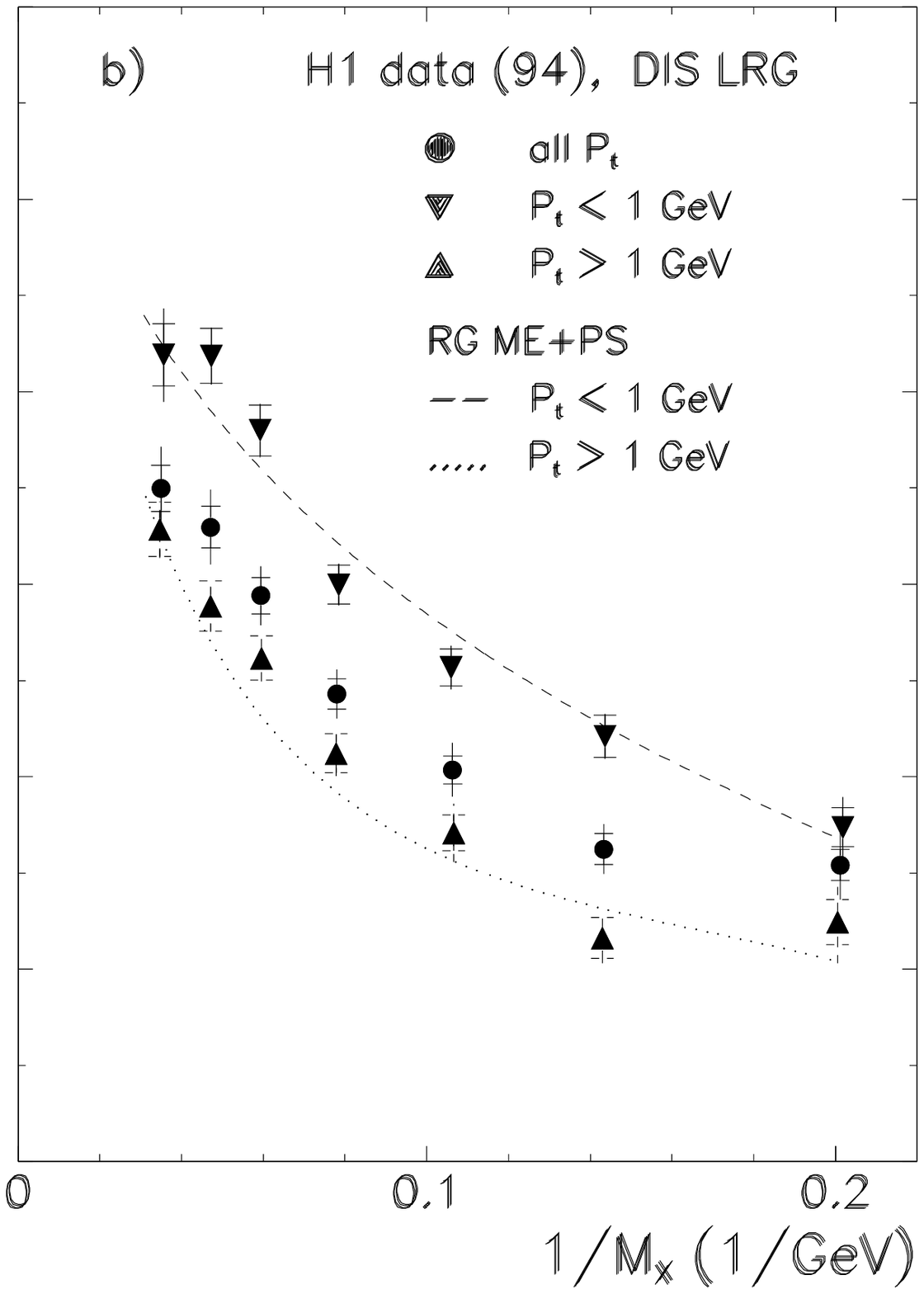,width=110mm,height=120mm}}
\end{picture}
%%    \parbox[t]{1cm}{~}
%%    \hspace{-.8cm}
%   \vspace{-.1cm}
    \parbox[t]{15.cm}
  {{\bf Figure 4a:} (a)~ \small                 
    Average thrust $\langle T\rangle $,
    as a function of $1/M_X$
    for LRG data, and for 
    $\ee$ annihilation data~\cite{PLUTO,EPEM_DAT} 
    as a function of $1/\sqrt{s_{ee}}$;
%%    \hspace{.7cm}
%   \vspace{3.1cm}
%%    \parbox[t]{7.cm}
    {\bf 4b:}~
    Average thrust $\langle T\rangle $,
    as a function of  $1/M_X$,  
    for all events,
    for events with $P_t <1$ GeV and for events with $P_t>1$ GeV,
    with predictions from RAPGAP (see text).
    The data points are given at $1/{\langle M_X\rangle} $.
}
\end{figure}
\begin{table}[ht]\centering
\begin{tabular}{|c|c|c|c|c|} \hline
$\langle M_X\rangle$(GeV)  & $\langle T\rangle $ & stat.error & syst.error &
B 
%  \\ \hline
%\multicolumn{4}{|c|} {all $P_t$}    
\\ \hline
%& \multicolumn{1-4}{|c|c|c|c|c|}
  4.97  &   0.777  &   0.004 &  0.008 & 0.18 \\
  6.98  &   0.781  &   0.004 & 0.005 & 0.16 \\ 
  9.40  &   0.802  &   0.004 & 0.006 & 0.18 \\
  12.81 &    0.822 &    0.004 & 0.005 & 0.16 \\
  16.82 &    0.847 &    0.005 & 0.006 & 0.12 \\
  21.20 &    0.865 &    0.005 & 0.008 & 0.10 \\
  28.64 &    0.875 &    0.006 & 0.009 & 0.10 
\\ \hline

\end{tabular}
\caption{\small Mean thrust values as a function of $\langle M_X\rangle$.
The following bin limits for $M_X$ were used:
4, 6, 8, 11, 15, 19, 24, 36 GeV. 
%The last column gives the relative weights with which results 
%from different mass intervals may be combined.}
To combine values of $\langle T\rangle $ from different $M_X$ intervals, the
relative cross section weights B are to be used.}
\label{table:thrust}
\end{table}
The correlation of thrust with $1/M_X$ rather than $M_X$ is shown
because it allows  
perturbative and hadronization effects on thrust to be distinguished
in a simple manner.
For $M_X \rightarrow \infty$ 
the latter should become negligible.
% and at finite $M_X$
%they should reduce $T$ compared to the parton level value 
%by an amount that is approximately proportional to $1/M_X$. 
%This is because the total momentum sum of the hard partons grows with $M_X$ 
%approximately linearly, but the 
%decorrelating transverse momenta coming from hadronization are 
%approximately $M_X$ independent.
%
For comparison, Fig.~4a includes  the available data for thrust 
in \ee annihilation events~\cite{PLUTO,EPEM_DAT} 
plotted at $M_X = \sqrt{s_{ee}}$.
%\medskip\\
The following features are noteworthy:
\medskip\\
\hspace*{0.5cm}
1. $\langle T\rangle $ increases with $M_X$. 
It is also observed in these data  that 
the particle multiplicity (not shown here) also increases with $M_X$.
This  implies that the final state is not isotropic,
since in that case $\langle T\rangle $ would fall, 
and approach 0.5, as $1/M_X \rightarrow 0$.  
The growth of $\langle T\rangle $ signifies
an increasing back-to-back correlation of
the momentum flow as the final state mass increases.
Hence, the thrust direction corresponds 
to an effective topological axis of the event.
This feature rules out fireball type models~\cite{FIRE_BALL}
with isotropic configurations of final state hadrons in
virtual photon diffractive dissociation.
Similar effects have also been
observed in hadronic single diffractive dissociation~\cite{UA45}.
\medskip\\
\hspace*{0.5cm}
2. There is a striking similarity between  the $\sqrt{s_{ee}}$
dependence of  $\langle T\rangle $ in \ee annihilation and 
its $M_X$ dependence in virtual photon dissociation,
for the kinematical range of this analysis.
%under the assumption that the role of $\sqrt{s_{ee}}$ in \ee annihilation 
%is played by $M_X$ in the DIS LRG events.
The slopes are comparable, but $\langle T\rangle $ of DIS LRG
events is lower by an approximately constant value of 0.025.
This similarity is not so evident in~\cite{ZEUSSPH}.
%\medskip\\

For $4 < M_X < 36$~GeV the thrust-mass correlation may be described by
\begin{equation}
 \langle T\rangle = T_{\infty}-H/M_X+F/M_X^2,
\end{equation}
as first observed in \ee annihilation experiments~\cite{PLUTO}.
There $T_{\infty}$ is related to thrust at the parton level and hence to 
\asx~\cite{RUJULA} in the $s_{ee}$ range of the data.
The coefficient $H$ describes the contribution to thrust of 
jet broadening  by hadronization.
Such a power correction term  was theoretically 
predicted~\cite{RUJULA},
confirmed by data~\cite{PLUTO},
reproduced with simulated events
using limited $p_t$ hadronization~\cite{FEYNMAN_FIELD} and is now 
discussed in the context of non-pertubative QCD~\cite{1OVERQ}.
%It is now the subject of theoretical considerations in the context 
%of renormalons~\cite{1OVERQ}.
Contrary to the situation in \ee annihilation, a theoretical justification of 
the $T_{\infty}$ and the $1/M_X$ terms  has not yet been given
for LRG events, but the data suggest that equ.(3)
applies here also.
The $1/M_X^2$ term is an ansatz to account for a bias of thrust 
 towards  large values when such a maximum search method is applied to
low particle
multiplicity final states which occur at low $M_X$.
In a fit of this parameterization to the $M_X$ range of the LRG data 
the parameter  $T_{\infty}$  is correlated  strongly with $H$ and $F$. 
To express the measured $\langle T\rangle$ dependence on $M_X$ 
with minimally correlated parameters an expansion of $\langle T\rangle$ 
in $1/M_X$ at a mass $M_X = M_0$ inside the data range  is appropriate:
\begin{equation}
\langle T\rangle = T_0-H\cdot (1/M_X-1/M_0)+F\cdot (1/M_X^2-1/M_0^2)
\end{equation}
Here $T_0$ is  $\langle T\rangle$ at $M_X = M_0$.
\begin{table}[ht]\centering
\begin{tabular}{|c|c|c|}                                           \hline
~~~~~~Data~~~~~~ &~~~~$T_0$~at $M_X=20$~GeV~~~~~~~~ & ~~~~~$H$ (GeV)~~~    \\ 
 \hline
\ep ~DIS LRG   & $ 0.857 \pm 0.003$       & $1.78 \pm 0.03$        \\
$\ee$ annihilation & $0.881 \pm 0.002$  &  $1.72 \pm 0.05$     \\  \hline
\end{tabular}
\caption{\small Results from the $T - M_X$ correlation (eqn.~2) fitted
 to this data and to \ee data~\cite{PLUTO,EPEM_DAT},
in the range from $6 < M_X < 36$~GeV.
$T_0$ is the  value of $\langle T\rangle$ at $M_X=20$~GeV, 
and $H$ the coefficient of the hadronization ($1/M_X$) term. 
The errors given are statistical errors for  the \ep ~data
and  statistical and systematic errors combined for the \ee data.
For the \ep ~data $T_0$ has in addition a systematic error of $\pm 0.006$.
The $H$ values for \ep ~and \ee have in addition a common error of 0.11 from 
their correlation to the uncertainty 
in the low multiplicity fluctuation term $F$.  
}
\label{table:TMXpars}
\end{table}
In a fit, with $M_0 = 20$~GeV,  
the coefficient $F$ for the $\ee$ annihilation data
was poorly constrained and was thus  fixed to the result   
found in LRG events ($F=5.0 \pm 0.6$~GeV$^2$).
The results of the fits in Table~\ref{table:TMXpars} show that
the power corrections ($H$) are compatible, 
while the thrust values are significantly lower in 
LRG than in \ee events. 
%The observation of comparable power corrections in \ee and LRG events
%contradicts the conclusions of Vermaseren et al.~\cite{VBLY}.
\medskip\\
\hspace*{0.5cm}
The result ${\langle T \rangle}_{LRG} < {\langle T \rangle}_{ee}$ 
means that the final state in DIS LRG events cannot be understood  
as a simple \qq  state 
with standard limited $p_t$ hadronization~\cite{FEYNMAN_FIELD}.
It implies that higher parton multiplicities are 
even more important in DIS LRG events than in \ee annihilation.
For a \qq parton configuration, 
${\langle T \rangle}_{LRG}$ can only be less than ${\langle T \rangle}_{ee}$ 
if the power corrections are stronger than in \eex annihilation data.
Then the value of $H$ would have to be larger than found in this fit.
A quantitative interpretation of the difference 
${\langle T \rangle}_{LRG} - {\langle T \rangle}_{ee}$ in the PD picture
in terms of contributions from the Born-term, 
and from the QCD-Compton and BGF subprocesses (see Fig.~2), requires  
a treatment of the non-perturbative diffractive vertex in these diagrams.
%or over the parton densities in the diffractive exchange in the DIS view.
This is not yet available.
In the IMF picture, 
the expressions for ${\langle T \rangle}_{LRG}$ are divergent.

Contributions from underlying configurations with three, as in
Fig.~2b, c and d, or more partons  
can also be studied using
the correlation between $P_t$ and thrust.
In Fig.~4b we show  $\langle T\rangle$ for events with $P_t$ 
smaller and larger than 1~GeV. 
 At each $M_X$ point average thrust for large $P_t$ is lower than for small $P_t$,
demonstrating the presence of 
configurations with at least 3 (non-colinear) partons.
The observed difference of $\langle T\rangle$ 
cannot be attributed to the selection of events with different
hadronization properties.  
If this were the case, the difference  would have to be proportional
to $1/M_X$, and this is not compatible with the data.
On the contrary, the difference is compatible with a higher,
$M_X$ independent, parton multiplicity.
Monte Carlo studies confirm that this $P_t$ dependent  $\langle T\rangle$
behaviour is that expected with 3-parton configurations,
as seen in the curves in Fig.~4b, 
which show  RAPGAP (ME+PS) expectations for 
a mixture of \qq and \qqg configurations.

%These predictions are similar for the ARIADNE mode, and
%they are almost independent
%of the $\hat p_{t,min}$ cut.
\par
The dependence of  $\langle T\rangle$ on $Q^2$ was also studied.
The value of  $\langle T\rangle$ changes by no more than
$\pm 0.01$ over the $Q^2$ range 10 to 100 GeV$^2$.
Since $\langle M_X\rangle$ is found to be independent of $Q^2$, 
it can be concluded that the data show almost no correlation between 
thrust and $Q^2$.
 
\subsection{${\bf P_t^2}$ distributions}
The cross sections differential in $P_t^2$
are shown in Fig.~5a and in Table~\ref{table:dN/dpt**2},
for six $M_X$ intervals in the mass range from 6 to 36 GeV, 
and normalized to unity in each $M_X$ interval. 
The $M_X$ interval from $4$ to $6$ GeV is not included since the 
flattening of $\langle T\rangle$ with $1/M_X$ in Fig.~4a, indicates 
that effects
from  low multiplicity fluctuations are becoming strong.
%
%--- Figure 5a/5b  : pt dependence
%
\begin{figure}[th] \unitlength 1mm
\label{figure:ptall}
\begin{picture}(160,100)
     \put(-5,0){\epsfig{file=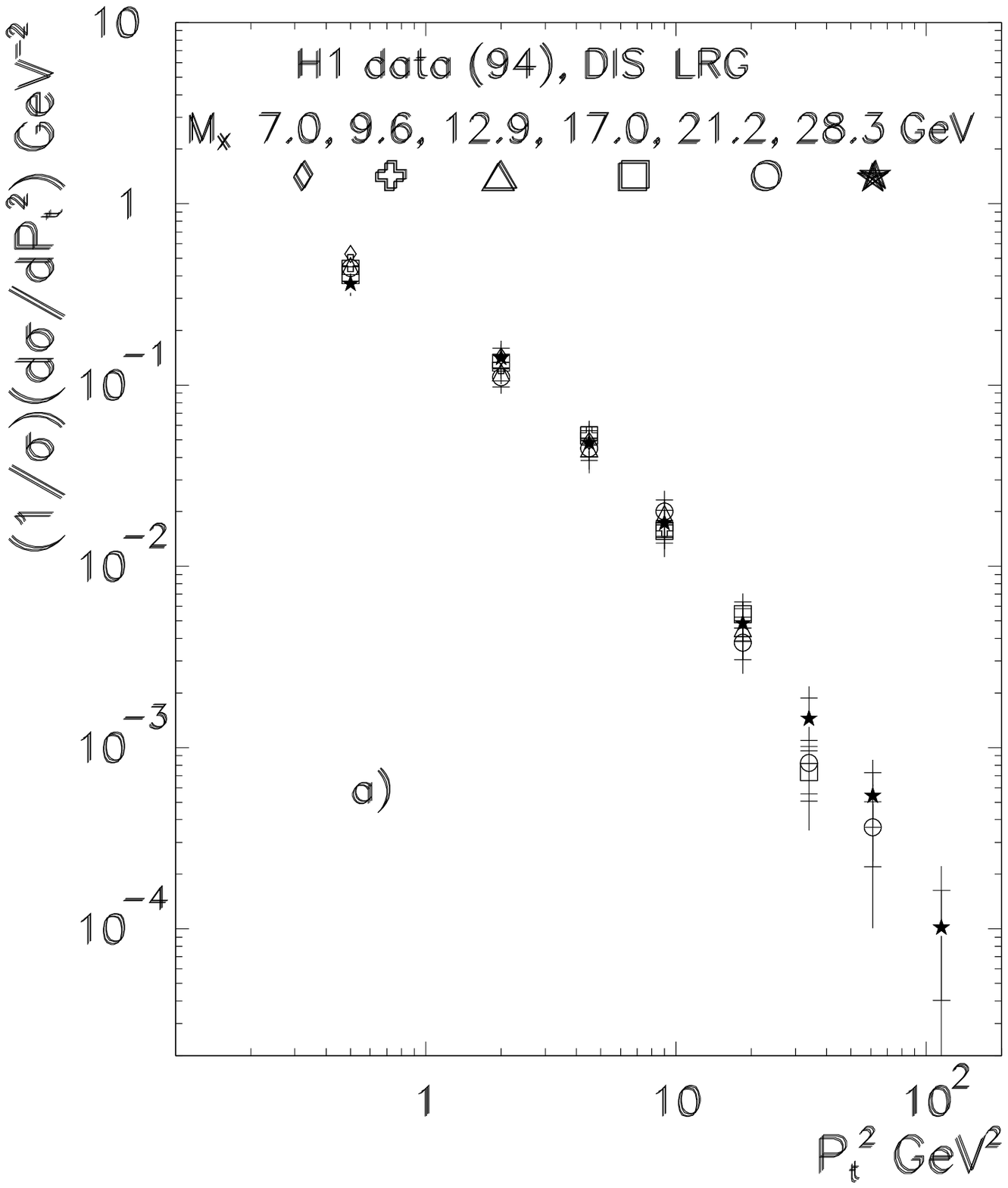,width=110mm,height=110mm}}
     \put(65,0){\epsfig{file=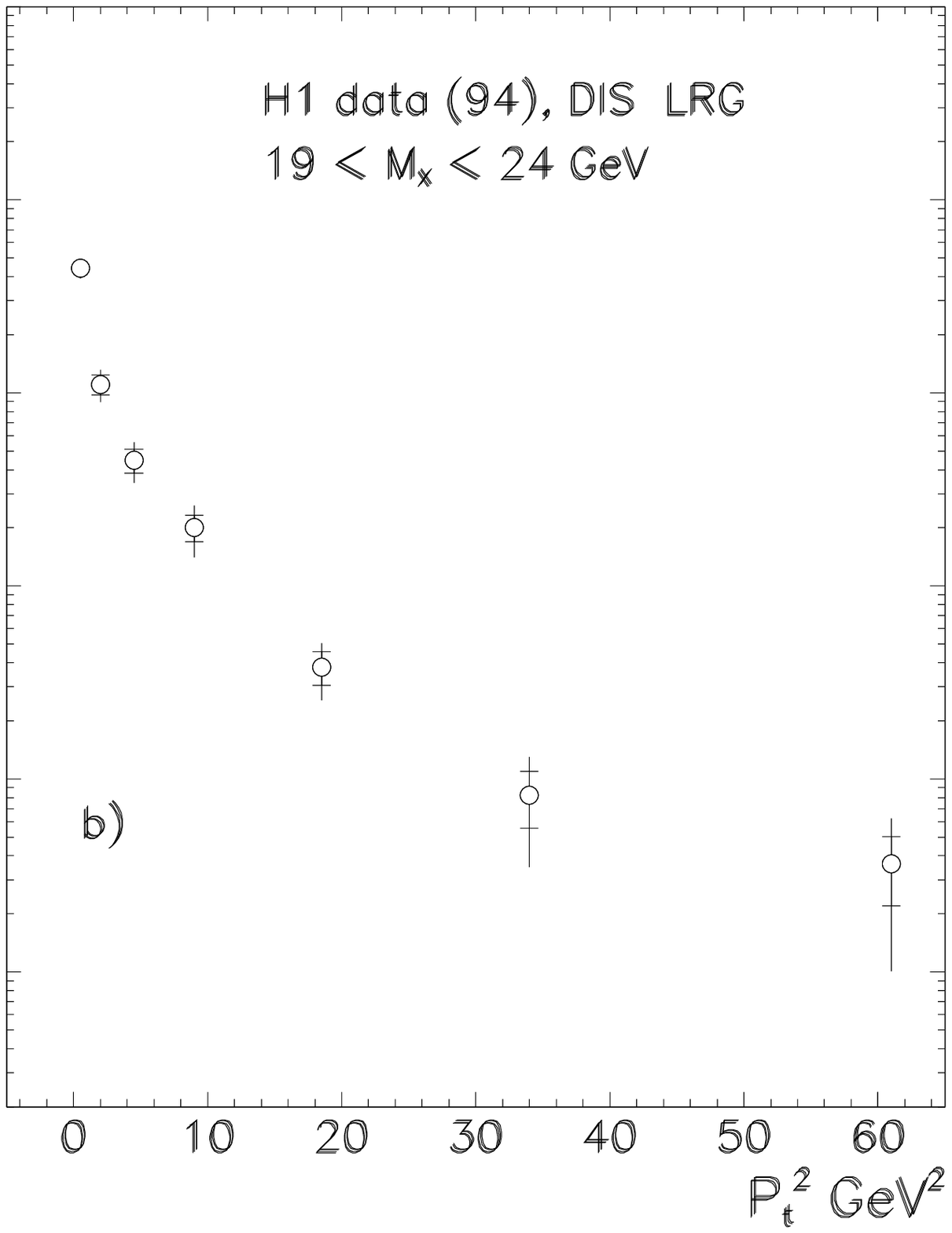,width=100mm,height=110mm}}
\end{picture}
%    \parbox[t]{1cm}{~}
%    \hspace{0cm}
     \parbox[t]{15.cm}
  {{\bf Figure 5a:}\small
    ~Normalized thrust jet $P_t^2$ distributions for six $M_X$ intervals. 
    The $P_t^2$ bin limits  are 0, 1, 3, 6, 12, 25, 43, 79,
    and 151 ${\rm GeV^2}$; 
%    \hspace{.5cm}
%    \parbox[t]{6.9cm}
   {\bf 5b:}                  
    Normalized thrust jet $P_t^2$ distribution for $19 < M_X<24$ GeV
    with a linear $P_t^2$ scale.    
    }
%\caption
%  {\small a:
%    Normalized thrust jet $P_t^2$ distributions for 6 $M_X$ intervals, 
%    The $P_t^2$ intervals are 0, 1, 3, 6, 12, 25, 43, 79,
%    and 151 $(GeV)^2$. 
% \newline
%    b:                    
%    Normalised thrust jet $P_t^2$ distributions for $19 < M_X < 24 GeV$.
%    The curves represent the 3 model predictions ( see text )
%}
\end{figure} 
\begin{table}[ht]\centering
\begin{tabular}{|c|c|c|c|c|} \hline
$\langle P_t^2\rangle $ GeV$^2$ & $\langle M_X\rangle $ (GeV) 
& $(1/\sigma)(d\sigma/dP_t^2)$ GeV$^{-2}$  & stat.error & syst.error
  \\ \hline
 0.37  & 6.96  &    0.527   &  0.018    &     0.038 \\
 1.88   & 6.96  &    0.144   &  0.009 &  0.022 \\
 4.21  & 7.10   &    0.042 & 0.004 &  0.007 \\
\hline
 0.38 & 9.39 &     0.470   &  0.017    &     0.039 \\
 1.84  &  9.36  &  0.129   &  0.008 &  0.022 \\
 4.30 &  9.31  &   0.052 &   0.004 &  0.007 \\
 8.38 &  9.54  &   0.016 &   0.002 &  0.003 \\
\hline
 0.40 &  12.72 &     0.451 &  0.019   &      0.033 \\
 1.83 &   12.74 &    0.117 &   0.009 &    0.013 \\
 4.22 &  12.77  &    0.045  & 0.004  &  0.005 \\
 8.54 &  12.78   &     0.019 &  0.002 & 0.003 \\
 15.97 &  13.10  &     0.0051 & 0.0007 &  0.0013 \\
\hline
  0.41 &  16.90 &     0.404 &  0.024   &      0.038 \\
 1.79  &   16.78 &    0.133 &  0.012 &   0.016 \\
 4.32  &   16.70  &     0.053 &  0.006 &  0.008 \\
 8.47  &  16.98   &     0.016 &  0.002 &  0.004 \\
 17.10 &   16.89  &     0.005 & 0.001  &  0.001 \\
 30.75 &   16.85  &    0.0007 &  0.0002 &  0.0003 \\
\hline
 0.37 &   21.01 &     0.444 &  0.030    &     0.044 \\
 1.84 &   21.10 &     0.111 & 0.013  &  0.016 \\
 4.22 &   21.40 &     0.045 &  0.006 &  0.009 \\
 8.08 &  20.91  &     0.020 & 0.003 &  0.005 \\
 16.19 &  21.07 &    0.0038 &  0.0008 &  0.001 \\
 33.18 &   20.79 &     0.0008 & 0.0003 &  0.0004 \\
 51.70 &  22.15  &       0.00039 & 0.00014 &  0.00022 \\
\hline
 0.41 &   27.79 &      0.361 &  0.032   &      0.039 \\
 1.83 &  28.05  &      0.141 & 0.018 & 0.028 \\
 4.26 &   27.94 &     0.048  & 0.008 &  0.013 \\
 8.42 &  28.26  &       0.017 & 0.003 &  0.004 \\
 17.38 &   28.65 &       0.0049 & 0.0010 &  0.0011 \\
 32.09 & 29.35    &      0.0014 & 0.0004 &  0.0006 \\
 58.28 &    28.18 &      0.00054 & 0.00018 &  0.00025 \\
 113.20 &   30.45 &    0.00011 & 0.00006 &  0.00011 \\
\hline
\end{tabular}
\caption{\small The $(1/\sigma)(d\sigma/dP_t^2)$ distribution. 
The following bin limits were used: 
0, 1, 3, 6, 12, 25, 43, 79, 151 GeV$^2$ for
$P_t^2$, and 6, 8, 11, 15, 19, 24, 36 GeV for $M_X$.
The table gives the mean values of $P_t^2$ and 
$M_X$ in each 2-dimensional bin, the differential
cross section in this bin divided by the total cross section for the
respective $M_X$ bin, and the statistical and the systematic errors
in the same normalized units.
The weight factors from Table~\ref{table:thrust} are to be applied
to  combine distributions from different mass bins.} 
\label{table:dN/dpt**2}
\end{table}
For each $M_X$ interval the data are displayed
only in those bins of $P_t^2$ that are
kinematically fully accessible.
In Fig.~5b the $P_t^2$ 
distribution of one mass range ($19 < M_X < 24$~GeV) is 
displayed on a linear scale in $P_t^2$. 
%
%\par
The data show:
\begin{enumerate}
\item a steep rise of the distribution towards low $P_t^2$ values, 
 demonstrating a dominant alignment of the final state thrust axis
 with the initial $\gamma^* p$ collision direction;
\item a shape of the $P_t^2$ distributions 
nearly independent of $M_X$,
suggesting the factorizable form
\begin{equation}
{\rm d}\sigma(M_X, P_t^2) /{\rm d}P_t^2 =A(M_X)\cdot B(P_t^2) ;
%1/\sigma\cdot{\rm d}\sigma(M_X, P_t^2) /{\rm d}P_t^2 =B(P_t^2) ;
\end{equation}
\item a substantial high $P_t^2$ tail.
\end{enumerate}

\par

%Since $P_t \propto M_X \cdot \sin\Theta$ the form (5) implies
%that the $\cos\Theta$ distributions of the thrust direction 
%are not independent of $M_X$.
%This is not compatible with the $M_X$ independent  $\cos\Theta$
%distributions reported by the ZEUS experiment~\cite{ZEUSSPH}.

The large high-$P_t$ tail supports the conclusion in Sect.~7.1
that there are significant contributions from parton multiplicities
higher than \qq,
such as those shown in Figs.~2b-d.
%The compatibility with a $P_t^{-4}$ dependence of the distributions in Fig.~5a 
%is - barring kinematic limit effects - in qualitative agreement with 
%the expectation for the hard subprocesses in those diagrams~\cite{NNNZAK}.  
The same conclusion may be gained from
Fig.~6 where  the fractions of events with  $P_t^2$ values 
above 1.0~${\rm {GeV}^2}$ and above 3.0~${\rm GeV^2}$, respectively,
are displayed as a function of $M_X$.
%
%--- Figure 6a/6b  : frractions and MC
%
\begin{figure}[th] \unitlength 1mm
\label{figure:mocas}
%%\begin{picture}(0,0)(0,0)
\begin{picture}(140,95)
     \put(-10,0){\epsfig{file=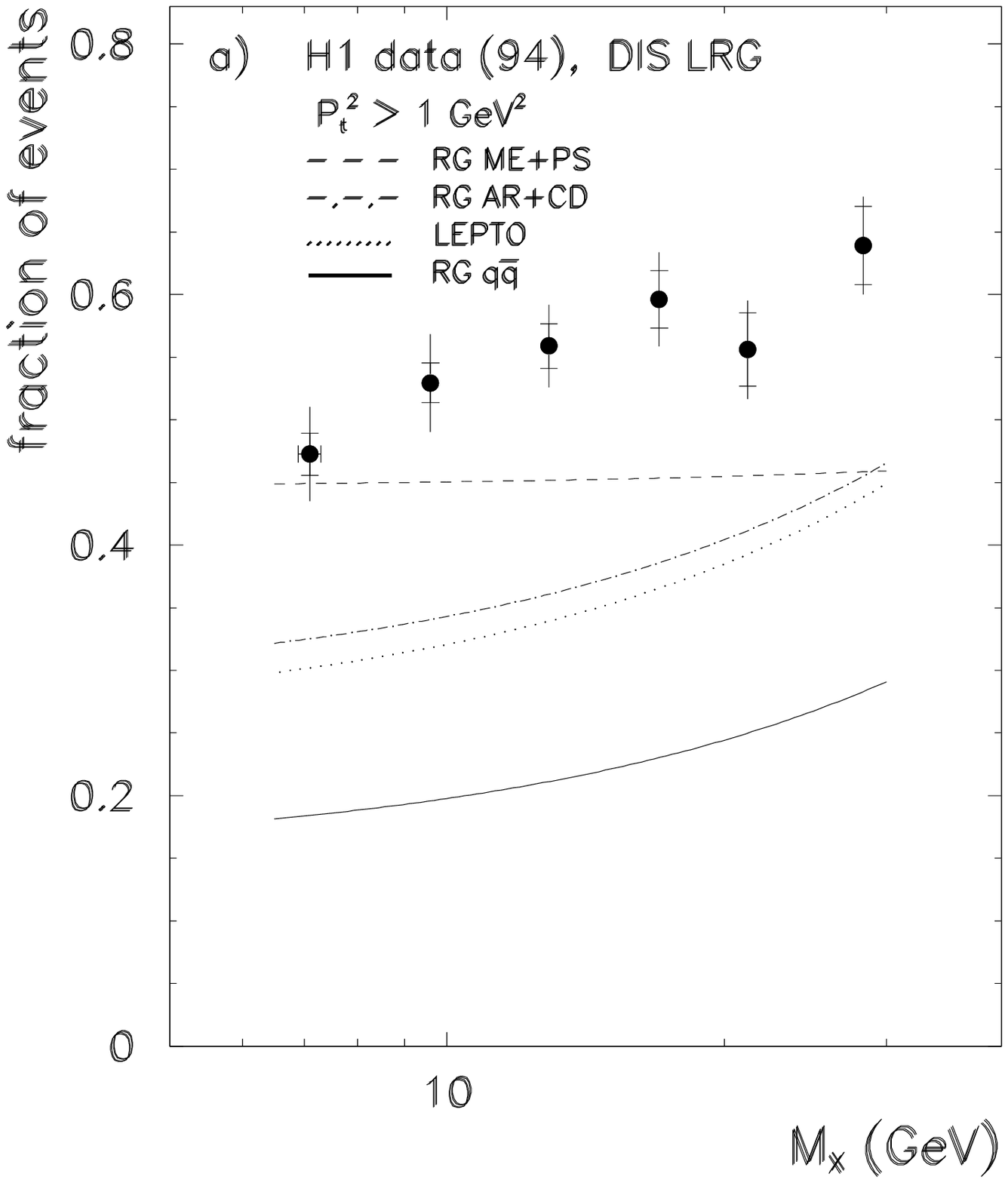,height=95mm,width=95mm}}
     \put(70,0){\epsfig{file=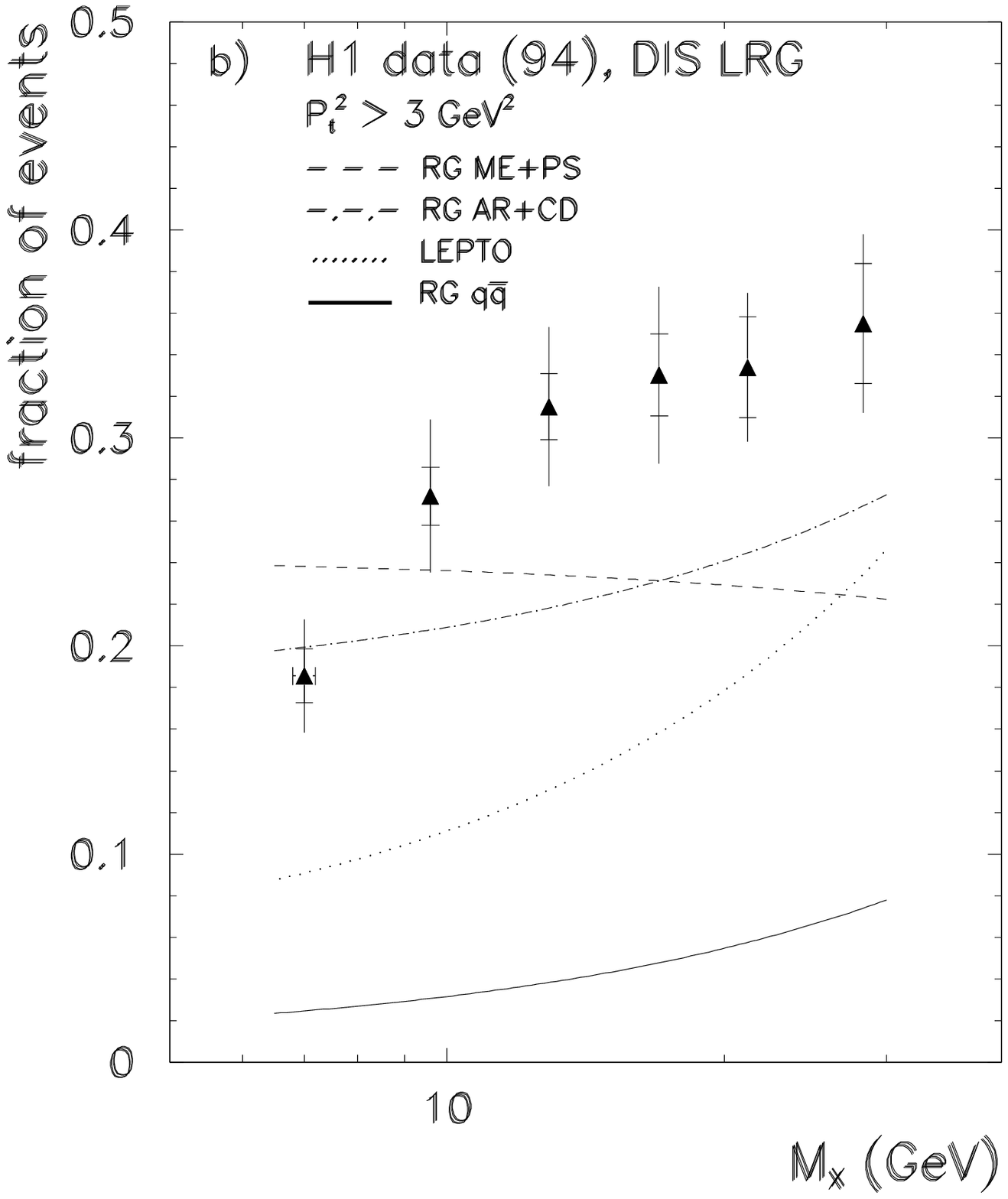,height=95mm,width=95mm}}
\end{picture}
%    \parbox[t]{1cm}{~}
%    \hspace{1cm}
%% This is the width of the text under the figure
    \parbox[t]{15.cm}
  {{\bf Figure 6a:} \small     
   Fraction of events with $P_t^2 >1~{\rm GeV^2}$ for six $M_X$ 
    intervals, together with four model predictions (see text); 
%   \hspace{.5cm}
%    \parbox[t]{7.5cm}
   {\bf 6b:}
  as in a, but with $P_t^2 >3~{\rm GeV^2}$.
}                   
\end{figure}
The fraction at the lowest $M_X$ point in Fig.~6b 
is relatively low because of the reduced phase space for
$P_t^2 > 3~{\rm {GeV}^2}$ here (see Table~\ref{table:dN/dpt**2}).
For higher masses the fractions
reach about 64\% for $P_t^2 > 1~{\rm {GeV}^2}$ and
35\% for $P_t^2 > 3~{\rm {GeV}^2}$.
The model predictions indicate that
such large values of $P_t^2$ require  hard processes, 
which involve more than two final state partons,
to play an important role in DIS LRG events.

%

%%%%%%%%%%%%%%%%%%%%%%%%%%%%%%%%%   Comparison to models %%%%%%%%%%%%%%%%%%%

\section{Comparison with models}

The GVDM model in combination with  quasi-elastic diffractive
vector meson proton scattering~\cite{SCHILDA} leads 
to final states  $X$ with spin $J=1$. For such a situation the
width of the $P_t^2$ distribution is proportional to $M_X^2$. 
Within present statistics the data do not support such a contribution.
Thus the spin parity structure of $X$ is neither pure nor dominantly vector.

%
%

%%%%%%%%%% replaced by Fig6,  end

%%--- Figure 6 T _ Mx correlation, comp with MC
%
\begin{figure}[h] \unitlength 1mm
%\label{figure:pthigh}
\begin{center}
\begin{picture}(140,100)
     \put(25,0)
{\epsfig{file=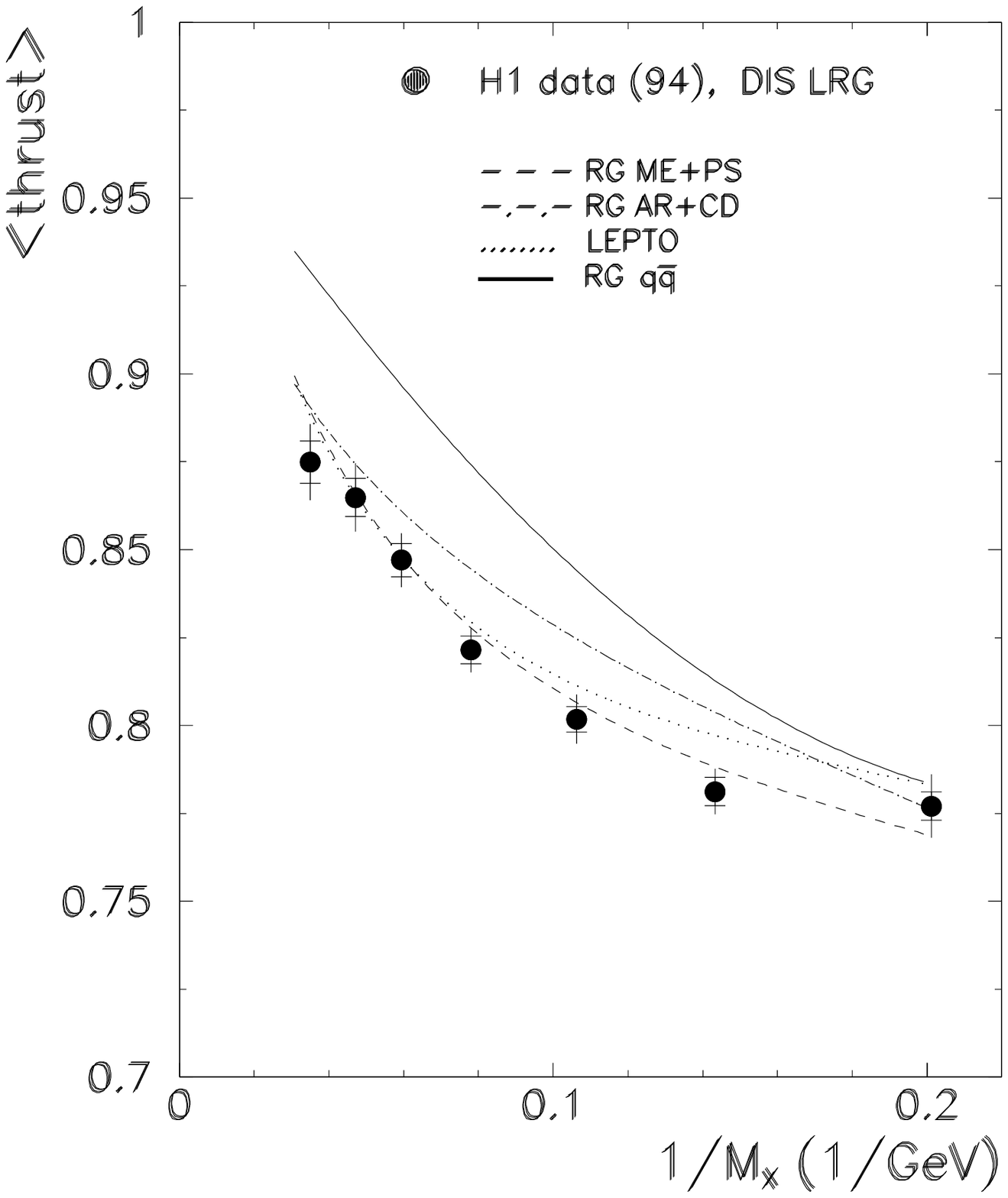,height=110mm,width=110mm}}
\end{picture}
    \parbox[t]{1cm}{~}
    \hspace{1cm}
%% This is the width of the text under the figure
    \parbox[t]{13.0cm}
  {{\bf Figure 7:}\small
    ~Comparison of model predictions (see text)
    with  the $M_X$ dependence of $\langle T \rangle$ .
  }
\end{center}
%    \hspace{.5cm}
%    \parbox[t]{6cm}
%  {Figure 5b:                  
%      Blablablabla}
\end{figure}

The  SCI model as presently implemented in LEPTO 
reproduces the $M_X$ dependence of 
$\langle T \rangle$ as shown in Fig.~7,
but fails to describe the fractions of events 
with $P_t^2$ larger than 1 or 3 GeV$^2$ (Fig.~6). 
This model as proposed has no means with which to adjust 
this prediction and therefore it cannot describe the LRG events exclusively.

The PD picture is qualitatively consistent
with two independent features of the data. 
Firstly, the strong, $M_X$ independent, alignment peak seen in Fig.~5
 is expected because of colour transparency.
%Secondly, the expected enhancement of the
%probability for  diffractive configurations with parton multiplicities larger
%than \qq,
%in particular for the BGF process,
%is supported both by the large $P_t^2$ tail
%and by  the observation that ${\langle T \rangle}_{LRG} < {\langle T \rangle}_{ee}$.
Secondly, the cross section ratio for \qqg to \qq configurations
is expected to be larger than in \ee annihilation,
mainly because of the large contribution for the BGF process,
again as a consequence of colour transparency. 
This is supported by the observation of
${\langle T \rangle}_{LRG} < {\langle T \rangle}_{ee}$ in Fig.~4a,
and by the large $P_t^2$ tail in Fig.~5. 
The qualitative expectations are confirmed by calculations 
in limited regions of phase space~\cite{BUCHJET}.

Calculations of the IMF picture 
in the RAPGAP implementation are shown in 
Figs.~6 and  7.
The AR+CD mode, which does not employ matrix elements
for the QCD-C process, 
 systematically predicts thrust too high and fractions
of events with large $P_t$ too low.
%This may indicate that the dipole radiation treatment for the QCD-C processes
%is insufficient to generate the proper momentum flow broadening.
The ME+PS predictions are closer
than those for the AR+CD mode, but the
agreement with the data is not completely satisfactory.
There are, however, several means to adjust the RAPGAP predictions to the data. 
One is a modification of the $\hat p_t$ cut-off scheme.
The present scheme, with an $M_X$ independent $\hat p_t$ cut-off value, 
is responsible for the counter-intuitive behaviour
of the prediction for the $M_X$ dependence of the large $P_t$
event fractions in the ME+PS mode.
In view of these model implementation problems,
one cannot conclude that this approach is unable to describe the data.

%%%%%%%%%%  replaced by Fig.~7, begin
%--- Figure 6a/6b  : frractions and MC
%
%\begin{figure}[ht] \unitlength 1mm
%\label{figure:mocas}
%%%\begin{picture}(0,0)(0,0)
%\begin{picture}(140,100)
%     \put(-10,-5){\epsfig{file=thrfig7a.ps,height=105mm,width=95mm}}
%     \put(70,0){\epsfig{file=thrfig7b.ps,height=100mm,width=95mm}}
%\end{picture}
%    \parbox[t]{1cm}{~}
%    \hspace{1cm}
%% This is the width of the text under the figure
%    \parbox[t]{7.5cm}
%  {Figure 6a:      
%  Thrust - $M_X$ correlation, comparison with 
%  4 model predictions (see text).}
%   \hspace{.5cm}
%    \parbox[t]{7.5cm}
%  {Figure 6b:                  
%     Fraction of events with $P_t^2 >3~{\rm (GeV)^2}$ for 6 $M_X$ 
%    intervals, together with 4 model predictions (see text). }
%\end{figure}

The RG q$\bar{\rm q}$ curves, 
which approximate the simple AJM, indicate the
extent to which the virtual photon dissociation process to \qq only is
unable to describe the data.

%%%%%%%%%%%%%%%%%%%%%%%% discussion %%%%%%%%%%%%%%%%%%%%%%%%%%%%%%%%%%%%%
%\newpage

\section{Summary and conclusions}

A thrust analysis of DIS large rapidity gap (LRG) events, 
that is of events attributable to the process $\gsp \rightarrow XY$, 
with $M_X > 4$~GeV,~$M_Y < 1.6$~GeV,~${\xpom}<0.05$
and~$10 < Q^2 < 100$~GeV$^2$ reveals:
\begin{itemize} 
\item
a dominant two-jet topology  of the final state $X$; 
\item
a clear alignment of the thrust axis with the incoming proton direction 
in the $X$ rest frame;
\item
a dependence of the average thrust value $\langle T\rangle $ 
on $P_t$, the thrust jet momentum transverse to the incoming proton direction: 
$\langle T\rangle $ decreases when  $P_t$ increases; 
\item
a $P_t^2$ spectrum independent of $M_X$;
\item
a sizeable fraction of events with large $P_t^2$ ($>3$~GeV$^2$);
\end{itemize}
and in comparison with \ee:
\begin{itemize}
\item
the influence of hadronization on $\langle T\rangle $  
is compatible with \ee;
\item
$\langle T\rangle $ is smaller than in \ee at $M_X = \sqrt{s_{ee}} $.
\end{itemize}

These features together cannot be explained by a pure \qq configuration. 
It has been shown in a model independent way that a substantial contribution
of \qqg and higher multiplicity parton configurations 
is required in the final state $X$.
This contribution is greater than in \ee annihilation data. 

There are  two equivalent pictures for diffractive LRG event production.
In the first the virtual photon fluctuates into multi-parton states 
together with quasi-elastic parton-proton diffraction.
In the second the electron scatters off partons which 
constitute the diffractive exchange.
The expectations of both pictures are broadly in agreement 
with the results of this analysis.
Two other mechanisms for LRG event production,
GVDM in the particular combination with 
quasi-elastic vector meson proton diffraction,
and non-diffractive \ep~DIS plus subsequent soft colour interaction, 
do not explain all topological properties of the observed events.

The abundance of parton multiplicities higher than \qq 
implied by topological event properties
is consistent with the large gluon content in diffractive exchange
found in QCD-fits to the inclusive production cross section 
of DIS LRG events~\cite{F2D3_new}.

%%%%%%%%%%%%%%%%%%%%%%%%Acknowledgement %%%%%%%%%%%%%%%%%%%%%%%%%%%%%%%%%%%%%
\section{Acknowledgements}
We are grateful to the HERA machine group whose outstanding efforts 
have made and continue to make this experiment possible.  

We thank the engineers and technicians for their work 
in constructing and now maintaining the H1 detector, 
our funding agencies for financial support,
%our love partners for their patience,
the DESY technical staff for continual assistance,
and the DESY directorate for the hospitality which they extend 
to the non-DESY members of the collaboration.
\vfill
\clearpage

%============ References =======================
%
%

\vfill
\clearpage
\newpage

\end{document}